\documentclass[aps,nofootinbib,showpacs,twocolumn]{revtex4}
\usepackage{graphicx}% Include figure files
\usepackage{dcolumn}% Align table columns on decimal point
\usepackage{bm}% bold math
\begin{document}

{LAPTH-1128/05, IFIC/05-60, APC-05-90}

\title{Probing neutrino masses with CMB lensing extraction}

\author{Julien Lesgourgues, Laurence Perotto}
\affiliation{Laboratoire de Physique Th\'eorique LAPTH
(CNRS-Universit\'e de Savoie), B.P. 110,
 F-74941 Annecy-le-Vieux Cedex, France}
\author{Sergio Pastor}
\affiliation{Instituto de F\'{\i}sica Corpuscular (CSIC-Universitat de
Val\`encia), Ed.\ Institutos de Investigaci\'on, Apdo.\ 22085,
E-46071 Valencia, Spain}
\author{Michel Piat}
\affiliation{Astroparticule et Cosmologie (APC),
Coll\`ege de France,
11 place Marcelin Berthelot, F-75231 Paris Cedex 05, France} 
\date{\today}

\begin{abstract}
We evaluate the ability of future cosmic microwave background (CMB)
experiments to measure the power spectrum of large scale structure
using quadratic estimators of the weak lensing deflection field.  We
calculate the sensitivity of upcoming CMB experiments such as BICEP,
QUaD, BRAIN, ClOVER and {\sc Planck} to the non-zero total neutrino
mass $M_\nu$ indicated by current neutrino oscillation data.  We find
that these experiments greatly benefit from lensing extraction
techniques, improving their one-sigma sensitivity to $M_{\nu}$ by a
factor of order four. The combination of data from {\sc Planck} and
the SAMPAN mini-satellite project would lead to $\sigma(M_{\nu}) \sim
0.1$ eV, while a value as small as $\sigma(M_{\nu}) \sim 0.035$ eV is
within the reach of a space mission based on bolometers with a
passively cooled 3-4 m aperture telescope, representative of the most
ambitious projects currently under investigation.  We show that our
results are robust not only considering possible difficulties in
subtracting astrophysical foregrounds from the primary CMB signal but
also when the minimal cosmological model ($\Lambda$ Mixed Dark Matter)
is generalized in order to include a possible scalar tilt running, a
constant equation of state parameter for the dark energy and/or extra
relativistic degrees of freedom.
\end{abstract}

\maketitle

\section{Introduction}

Nowadays there exist compelling evidences for flavor neutrino
oscillations from a variety of experimental data, that includes
measurements of solar, atmospheric, reactor and accelerator neutrinos
(for recent reviews, see e.g.\ \cite{Maltoni:2004ei,Fogli:2005cq}).
The existence of flavor change implies that the three neutrinos mix
and have non-zero masses, but oscillation experiments only fix the
differences of squared neutrino masses $\Delta m^2_{31}$ and $\Delta
m^2_{21}$, which correspond to the values relevant for atmospheric
($2.4\times 10^{-3}$ eV$^2$) and solar ($8\times 10^{-5}$ eV$^2$)
neutrinos, respectively.

Non-zero neutrino masses imply that the Cosmic Neutrino Background
(CNB), the sea of relic neutrinos that fill the Universe with a number
density comparable to that of photons, influences the cosmological
evolution in a more complicated way than that of a pure relativistic
component. In particular, the contribution of the CNB to the present
energy density of the Universe, measured in units of its critical
value, is
\begin{equation}
\Omega_{\nu} = \frac{\rho_\nu}{\rho_c} =
\frac{M_{\nu}}{93.14\,h^2~{\rm eV}}
\label{omega_nu}
\end{equation}
where $h$ is the present value of the Hubble parameter in units of
$100$ km s$^{-1}$ Mpc$^{-1}$ and $M_{\nu}\equiv m_1+m_2+m_3$ is the
total neutrino mass. {}From the experimental values of their mass
differences, at least two neutrino mass states are non-relativistic
today since both $(\Delta m^2_{31})^{1/2}\sim 0.05 $ eV and $(\Delta
m^2_{21})^{1/2}\sim 0.009$ eV are larger than the present neutrino
temperature $T_\nu \simeq 1.96$ K $\simeq 1.7\times 10^ {-4}$ eV.
Since the current upper bound on $M_\nu$ from tritium decay
experiments \cite{Eitel:2005hg} is of the order $6$ eV (95\% CL), we
know that the neutrinos account for at least 0.5(1)\% and at most 50\%
of the total dark matter density, where the lower limit corresponds to
the minimum of $M_{\nu}$ for masses ordered according to a normal
(inverted) hierarchy, characterized by the sign of $\Delta
m^2_{31}$. Thus, although in the first limit the cosmological effect
of neutrino masses would be quite small, the minimal cosmological
scenario is in fact a $\Lambda$ Mixed Dark Matter ($\Lambda$MDM) model
rather than a plain $\Lambda$ Cold Dark Matter one.

Considerable efforts are devoted to the determination of the absolute
neutrino mass scale, which, combined with oscillation data, would fix
the value of the lightest neutrino mass. The future tritium decay
experiment KATRIN \cite{Osipowicz:2001sq} is expected to reach a
discovery potential for $0.3-0.35$ eV individual masses, while more
stringent bounds exist from experiments searching for neutrinoless
double beta decay\footnote{A claim of a positive signal exists
\cite{Klapdor-Kleingrothaus:2004wj}, which would correspond to an
effective neutrino mass of order $0.1-0.9$ eV. If confirmed, it would
have a profound impact on cosmology.}. These will be improved in the
near future \cite{Elliott:2002xe}, but unfortunately they depend on
the details of the neutrino mixing matrix. The quest for $M_{\nu}$
will greatly benefit from cosmological observations, which offer the
advantage of being independent of the neutrino mixing parameters since
all flavors were equally populated in the early Universe.

Cosmology is sensitive to the neutrino masses through essentially two
effects.  First, the shape of the two-point correlation function --or
power spectrum-- of the Cosmic Microwave Background (CMB) temperature
and polarization anisotropies on the one hand, and of the Large Scale
Structure (LSS) mass density on the other, are both highly sensitive
to the abundance of the various cosmological backgrounds: photons,
baryons, cold dark matter, etc. The CNB is very specific in the sense
that it behaves like a collisionless relativistic medium at the time
of acoustic oscillations before photon decoupling (at redshifts $z >
1000$), but like a non-relativistic fluid during most of structure
formation (at redshifts $z < 100$, at least for one of the three
neutrino mass states). Therefore, the CNB affects at least one of the
three following quantities: the redshift of equality between matter
and radiation; the redshift of equality between matter and dark
energy; or the spatial curvature of the Universe. This effect can be
observed in the CMB and LSS power spectra and its amplitude is at most
of the order of $(2 f_{\nu})$ per cent \cite{review} ($f_{\nu}$ is the
current fraction of dark matter density in the form of neutrinos),
which corresponds to only 1\% in the limit $M_{\nu}\sim 0.05$ eV.

Fortunately, neutrino masses produce a second effect which is
typically four times larger: on small scales neutrinos do not cluster
gravitationally because of their large velocities. Even today, the
typical neutrino velocity of a non-relativistic eigenstate with mass
$m_{\nu}$ is as large as $v\simeq 150 \, (1 \, {\rm eV}/m_{\nu})$
km s$^{-1}$.  This simple kinematic effect, called {\it neutrino
free-streaming}, is extremely important for the growth of
non-relativistic matter perturbations (CDM and baryons) after photon
decoupling.  Indeed, the perturbation growth rate is controlled by the
balance between gravitational clustering and the Universe expansion.
On small scales, free-streaming neutrinos contribute to the total
background density $\bar{\rho}$, but not to the total perturbation
$\delta \rho$, which shifts the balance in favor of the Universe
expansion, leading to a smaller growth rate for CDM and baryon
perturbations. This effect is of order $(8 f_{\nu})$ per cent in the
small-scale matter power spectrum \cite{Bond:1980ha,Hu:1997mj,review}.

There are various ways to measure the LSS power spectrum. For
instance, the galaxy-galaxy correlation function can be obtained from
galaxy redshift surveys, and the density perturbations in hot
intergalactic gas clouds at redshift $z\sim2$ can be inferred from the
Lyman-$\alpha$ forest region in the spectrum of distant quasars. At
present, a total neutrino mass of $0.4-2$ eV is disfavored at 95\% CL
\cite{Spergel:2003cb,Hannestad:2003xv,Tegmark:2003ud,Barger:2003vs,Hannestad:2003ye,Crotty:2004gm,Rebolo:2004vp,Seljak:2004xh,Fogli:2004as,Ichikawa:2004zi,MacTavish:2005yk,Sanchez:2005pi},
depending on the used CMB, LSS and other cosmological data.

However, the most promising idea on the long term is to study
the weak lensing effects induced by neighboring galaxy clusters.  A
lensing map can be reconstructed from a statistical analysis, based
either on the ellipticity of remote galaxies or on the non-gaussianity
of the CMB temperature and polarization anisotropy maps. Weak lensing
offers several advantages. Unlike galaxy redshift surveys, it traces
directly the total density perturbation and does not involve any
light-to-mass bias. Unlike Lyman-$\alpha$ forests data, it probes a
large range of scales, which is particularly convenient for observing
the step-like suppression of density perturbations induced by neutrino
masses. In addition, weak lensing is sensitive to high redshifts, for
which non-linear corrections appear only at very small
scales. Finally, it enables tomographic reconstruction: by selecting
the redshift of the sources, it is possible to obtain independent
measurements of the power spectrum at various redshifts, in order to
follow the non-trivial evolution of the spectrum amplitude caused by
neutrino masses and/or by a possible evolution of the dark energy
density.  The best lever arm and the highest redshifts are encoded in
the lensing of CMB maps, where the source is the photon last
scattering surface, located at $z\sim 1100$, and the observed CMB
patterns are sensitive to lenses as far as $z\sim 3$
\cite{Ber97,LE,LE2}.  In addition, CMB lensing observations do not
require a devoted experiment: future CMB experiments designed for
precision measurements of the primary CMB anisotropies offer for free
an opportunity to extract lensing information.

The first paper estimating the sensitivity of future cosmological
experiments to small neutrino masses was based on the measurement of
the galaxy-galaxy correlation function \cite{Hu:1997mj}, an analysis
that was updated in Refs.\
\cite{Eisenstein:1998hr,Lesgourgues:1999ej,Hannestad:2002cn} and more
recently in Ref.\ \cite{Lesgourgues:2004ps}.  The idea that weak
lensing observations (from galaxy ellipticity) were particularly
useful for measuring the neutrino mass was initially proposed in Ref.\
\cite{Cooray:1999rv}. Then, the first analysis based on CMB lensing
extraction was performed in Ref.\ \cite{Kaplinghat:2003bh}, showing
that an extremely small one-sigma error on the total neutrino mass
--of the order of $\sigma(M_{\nu}) \simeq 0.04$ eV-- was conceivable
for a full-sky experiment with a resolution of 1 arc-minute and a
sensitivity per pixel of 1 $\mu$K for temperature, 1.4 $\mu$K for
polarization (these numbers were inspired from preliminary studies for
the CMBpol satellite project).  Soon after, Ref.\ \cite{Song:2003gg}
studied the neutrino mass sensitivity of future tomographic
reconstructions using, on the one hand, galaxy ellipticities in
various redshift bins, and on the other CMB lensing, where CMB plays
the role of the last redshift bin. The authors found that for
sufficiently large cosmic shear surveys, it would not be impossible to
reach $\sigma(M_{\nu}) \simeq 0.02$ eV.

In this paper we want to come back to the prospects coming from CMB
lensing alone, and try to improve the pioneering analysis in
\cite{Kaplinghat:2003bh,Song:2003gg} in several directions. First, we
analyze the potential of several CMB experiments expected to produce
results in the coming years, based on a realistic description of
instrumental sensitivities. Second, we discuss the robustness of our
results by analyzing (i) the consequences of simplifying assumptions
in the construction of the Fisher matrix, (ii) the dependence of the
final results on the accuracy of the foreground subtraction process,
and (iii) the impact of parameter degeneracies which can appear when
non-minimal cosmological scenarios are introduced.  Finally, we study
the sensitivity of CMB experiments to the way in which the total
neutrino mass is split among the three species.

\section{Basic principles of CMB lensing extraction}

Weak lensing induces a deflection field ${\bf d}$, i.e.\ a mapping
between the direction of a given point on the last scattering surface
and the direction in which we observe it. At leading order \cite{IE}
this deflection field can be written as the gradient of a lensing
potential, ${\bf d}=\nabla \phi$. The (curl-free) deflection map and
the lensing potential map can both be expanded in harmonic space
\begin{eqnarray}
\phi(\hat{n}) 
&=& 
\sum_{lm} \phi_{l}^{m} Y_l^m(\hat{n})~,\\
( d_{\theta} \pm i d_{\varphi} ) (\hat{n}) 
&=& 
\pm i \sum_{lm} d_{l}^{m\pm1} Y_l^m(\hat{n})~,
\end{eqnarray}
where $\hat{n}=(\theta, \phi)$ is a direction in the sky.
There is a simple relation between the deflection and lensing
potential multipoles
\begin{equation}
 d_{l}^{m} = -i \sqrt{l(l+1)} \phi_{l}^{m}~,
\end{equation}
so that the power spectra 
$C_l^{dd} \equiv \langle d_{l}^{m} d_{l}^{m*} \rangle $ and
$C_l^{\phi \phi} \equiv \langle \phi_{l}^{m} \phi_{l}^{m*} \rangle $
are related through
\begin{equation}
C_l^{dd} = l(l+1) C_l^{\phi \phi}~.
\end{equation}
In standard inflationary cosmology, the unlensed anisotropies obey
Gaussian statistics in excellent approximation \cite{NG}, and their
two-dimensional Fourier modes are fully described by the power spectra
$\tilde{C}^{ab}_l$ where $a$ and $b$ belong to the $\{T, E, B\}$
basis. Weak lensing correlates the lensed multipoles
\cite{Sel96,Ber97} according to
\begin{equation}
\langle a_l^m b_{l'}^{m'} \rangle_{\rm CMB}
= (-1)^m \delta_l^{l'} \delta_m^{m'} C^{ab}_{l}
+ \sum_{LM} {\cal C}(a,b)_{l \,\, l' \,\, L}^{mm'M} \,\, \phi_L^M
\label{ab}
\end{equation}
where the average holds over different realizations (or different
Hubble patches) of a given cosmological model with fixed primordial
spectrum and background evolution (i.e.\ fixed cosmological
parameters). In this average, the lensing potential is also kept fixed
by convention, which makes sense because the CMB anisotropies and LSS
that we observe in our past light-cone are statistically independent,
at least as long as we neglect the integrated Sachs-Wolfe effect.  In
the above equation, $C^{ab}_{l}$ is the lensed power spectrum (which
is nearly equal to the unlensed one, excepted for the B-mode power
spectrum which is dominated, at least on small scales, by the
conversion of E-patterns into B-patterns caused by lensing).  The
coefficients ${\cal C}(a,b)_{l \,\, l' \,\, L}^{mm'M}$ are complicated
linear combinations of the unlensed power spectra
$\tilde{C}^{ab}_{l}$, $\tilde{C}^{aa}_{l}$ and $\tilde{C}^{bb}_{l}$,
given in \cite{QE3}.

The quadratic estimator method of Hu \& Okamoto \cite{QE1,QE2,QE3} is
a way to extract the deflection field map from the observed
temperature and polarization maps.  It amounts essentially in
inverting Eq.~(\ref{ab}). This is not the only way to proceed: Hirata
\& Seljak proposed an iterative estimator method \cite{IE} which was
shown to be optimal, but as long as CMB experiments will make
noise-dominated measurements of the B-mode, i.e.\ at least for the
next decade, the two methods are known to be equivalent in terms of
precision. Even for the most precise experimental project discussed in
this work, the
quadratic estimator method would remain nearly optimal
(the last project listed in Table \ref{tableexp} corresponds 
roughly to the hypothetical experiment called ``C'' in Ref.\ \cite{IE}).

By definition, the quadratic estimator $d(a,b)$ is built from a pair
$(a,b)$ of observed temperature or polarization modes, and its
multipoles are given by the quadratic form
\begin{equation}
d(a,b)_L^M = {\cal N}^{ab}_L \sum_{ll'mm'} {\cal W}(a,b)_{l\,\,l'\,\,L}^{mm'M}
\,\, a_l^m b_{l'}^{m'}~,
\label{defd}
\end{equation}
where the normalization factor ${\cal N}_L^{ab}$ is defined in such way that
$d(a,b)$ is an unbiased estimator of the deflection field
\begin{equation}
\langle d(a,b)_L^M \rangle_{\rm CMB} = \sqrt{l(l+1)} \phi_L^M~,
\end{equation}
and the weighting coefficients ${\cal W}(a,b)_{l\,\,L\,\,l'}^{mm'M}$
minimize the variance of $d(a,b)_L^M$ (which inevitably exceeds the
power spectrum $C_L^{dd}$ that we want to measure), i.e.\ minimize the
coefficients $a=a'$, $b=b'$ of the covariance matrix
\begin{equation}
\langle d(a,b)_L^M d(a',b')_{L'}^{M'} \rangle_{\rm CMB} = 
(-1)^M \delta_L^{L'} \delta_M^{M'} [C^{dd}_{L} + N^{aba'b'}_L]~.
\end{equation}
Here the extra term $N^{aba'b'}_L$, which can be considered as noise,
derives from the connected and non-connected pieces of the 
four-point correlation function $\langle aba'b' \rangle$.
In Ref.~\cite{QE3}, Okamoto \& Hu derive a prescription for 
the weighting coefficients ${\cal W}(a,b)_{l\,\,L\,\,l'}^{mm'M}$
such that the contribution of the connected piece is minimal, while
that from the non-connected piece is negligible in first approximation
\cite{Coo02}.
The weighting coefficients are rational functions of the observed
power spectra $C_l^{ab}$, $C_l^{aa}$ and $C_l^{bb}$, 
which include contributions from primary 
anisotropies, lensing and experimental noise. Therefore, if we
assume a theoretical model and some instrumental characteristics, we
can readily estimate the noise $N^{aba'b'}_L$ expected for a future
experiment.

This method works for a given estimator $d(a,b)_L^M$ under the condition
that for at least one of the three power spectra ($C_l^{ab}$, $C_l^{aa}$,
$C_l^{bb}$), the lensing contribution is much
smaller than the primary anisotropy contribution.
This is not the case for the pair $ab=BB$. 
Therefore, one can only build
five estimators, for the remaining 
pairs $ab \in \{ TT, EE, TE, TB, EB \}$.
The question of which one is the most precise heavily depends on
the experimental characteristics. In addition, it is always possible
to build a minimum variance estimator, i.e.\ an optimal combination
of the five estimators weighted according to the five noise
terms $N^{aba'b'}_l$ of the experiment under consideration.
For the minimum variance estimator, the noise reads
\begin{equation}
N^{dd}_l = \left[ \sum_{aba'b'} \left( N^{aba'b'}_l \right)^{-1} \right]^{-1}.
\end{equation}

\section{Forecasting errors with the Fisher matrix}

For a future experiment with known specifications, it is possible to
assume a cosmological {\it fiducial model} that will fit best the
future data, and then to construct the probability
$L(\vec{x};\vec{\theta})$ of the data $\vec{x}$ given the parameters
$\vec{\theta}$ of the theoretical model.  The error associated with
each parameter $\theta_i$ can be derived from the Fisher matrix
\begin{equation}
F_{ij}=- \left\langle 
\frac{\partial^2 \ln L}{\partial \theta_i \partial \theta_j} 
\right\rangle~,
\end{equation}
computed in the vicinity of the best-fit model. Indeed, after
marginalization over all other free parameters, the one-sigma error
(68\% confidence limit) on a parameter $\theta_i$ would be greater or
equal to
\begin{equation}
\sigma(\theta_i) = \sqrt{(F^{-1})_{ii}}~.
\end{equation}
In most cases, the forecasted errors depend only
mildly on the exact values of fiducial model parameters; however,
they can vary significantly with the number of free parameters
to be marginalized out, since complicated fiducial models with many
physical ingredients are more affected by parameter degeneracies. 

It is usually assumed that for a CMB experiment covering a fraction
$f_{\rm sky}$ of the full sky, the probability $L$ of the data
$\{a_{l}^{m}\}$ is gaussian, with variance $\mathbf{C}_l$. If the
experiment observes only one mode, for instance temperature,
then $\mathbf{C}_l$ is just a number,
equal to the sum of the fiducial model primordial spectrum and of the
instrumental noise power spectrum. If instead several modes are
observed, for instance temperature, E and B polarization, then
$\mathbf{C}_l$ is a matrix. Neglecting the lensing effect, we would
get
\begin{equation}
\mathbf{C}_l=
\left( \begin{array}{ccc}
\tilde{C}_l^{TT}+N_l^{TT} & \tilde{C}_l^{TE} & 0\\
\tilde{C}_l^{TE} & \tilde{C}_l^{EE}+N_l^{EE} &0 \\
0 & 0 & \tilde{C}_l^{BB}+N_l^{BB} \\
\end{array}\right)~,
\label{covmat_nolens}
\end{equation}
where the $\tilde{C}_l^{XX}$'s represent the power spectra of primary
anisotropies (we recall that for parity reasons
$\tilde{C}_l^{TB}=\tilde{C}_l^{EB}=0$), and the $N_l^{XX}$'s are the noise
power spectra, which are diagonal because the noise contributing to
one mode is statistically independent of that in another
mode. It can be shown with some algebra that for any gaussian
probability $L$, the Fisher matrix
reads \cite{Tegmark:1996bz}
\begin{equation}
F_{ij} = \frac{1}{2} \sum_{l} (2l+1)f_{\rm sky}
\mathrm{Trace}[\mathbf{C^{-1}} \frac{\partial \mathbf{C}}{\partial \theta_i}
\mathbf{C^{-1}} \frac{\partial \mathbf{C}}{\partial \theta_j}]~.
\label{trace}
\end{equation}
In fact, due to the lensing effect, the data is not exactly
gaussian. However, the difference between the unlensed and lensed
power spectra for ($TT$, $TE$, $EE$) is so small that Eq.~(\ref{trace})
remains approximately correct, at least when the B-mode is not
included in the covariance matrix of Eq.~(\ref{covmat_nolens}). Beyond this
issue, lensing offers the possibility to include an extra piece of
information: namely, the map of the lensing potential --or
equivalently, of the deflection vector-- as obtained from e.g. the
quadratic estimators method. Ideally, after lensing extraction, one
would obtain four gaussian independent variables: the {\it delensed}
temperature and anisotropy multipoles ($\tilde{T}_l^m$,
$\tilde{E}_l^m$, $\tilde{B}_l^m$), and the lensing multipoles
$d_l^m$. In this paper, we will take a fiducial model with no
significant amplitude of primordial gravitational waves. In this case,
the delensed B-mode is just noise and can be omitted from the Fisher
matrix computation. Therefore the data covariance matrix reads
\begin{equation}
\mathbf{C}_l=
\left( \begin{array}{ccc}
\tilde{C}_l^{TT}+N_l^{TT} & \tilde{C}_l^{TE} & C_l^{Td}\\
\tilde{C}_l^{TE} & \tilde{C}_l^{EE}+N_l^{EE} &0 \\
C_l^{Td} & 0 & C_l^{dd}+N_l^{dd} \\
\end{array}\right)~,
\label{lensed_covmat}
\end{equation}
where $C_l^{dd}$ is the lensing power spectrum, $N_l^{dd}$ the noise
associated to the lensing extraction method (in our case, the minimum
variance quadratic estimator), and $C_l^{Td}$ the cross-correlation
between the unlensed temperature map and lensing map. This term does
not vanish because of the late integrated Sachs-Wolfe effect: the
temperature includes some information on the same neighboring cluster
distribution as the lensing. Both $C_l^{dd}$ and $C_l^{Td}$ can be
computed numerically for a given theoretical model using a public
Boltzmann code like {\sc camb} \cite{Lewis:1999bs}, and then
$N_l^{dd}$ can be calculated using the procedure of Ref.\
\cite{QE3}. This computation can be performed in the full sky: in this
work, we will never employ the flat-sky approximation.  Note that the
$B$-mode does not appear explicitly in Eq.\ (\ref{lensed_covmat}), but
actually information from the observed B-mode is employed in the two
estimators $d(T,B)$ and $d(E,B)$.

Using Eqs.\ (\ref{trace}) and (\ref{lensed_covmat}), it is possible to
compute a Fisher matrix and to forecast the error on each cosmological
parameter. Let us discuss the robustness of this method. There are obviously
two caveats which could lead to underestimating the errors.

First, we assumed in Eq.\ (\ref{lensed_covmat}) that the temperature
and polarization maps could be delensed in a perfect way. Instead, the
delensing process would necessarily leave some residuals, in the form of
extra power and correlations in the covariance matrix. However,
this is not a relevant issue, because we are using only
the temperature and $E$-polarization modes, for which the lensing
corrections are very small: therefore, considering a small residual or no
residual at all makes no difference in practice. We checked this
explicitly in a simple way. For a given theoretical model, Boltzmann
codes like {\sc camb} \cite{Lewis:1999bs}
are able to compute both the lensed and unlensed power
spectra. If the delensing process is totally inefficient, we can say
that unlensed temperature and polarization multipoles are recovered
with an error of variance
\begin{equation}
E_l^{TT}= | C_l^{TT} - \tilde{C}_l^{TT} | ~,
\qquad
E_l^{EE}= | C_l^{EE} - \tilde{C}_l^{EE} | ~,
\end{equation}
that we can treat as additional noise and sum up to the $N_l^{TT}$ and
$N_l^{EE}$ terms in the matrix (\ref{lensed_covmat}). We checked
numerically that even with such a pessimistic assumption, the final
result does not change significantly, which is not a surprise since
$E_l^{aa} \ll \tilde{C}_l^{aa}$. We conclude that the assumption of
perfect delensing performed in Eqs.~(\ref{trace}) and
(\ref{lensed_covmat}) is not a problem in practice\footnote{Note that
replacing $\tilde{C}_l^{TT}$ by $C_l^{TT}$ in (\ref{lensed_covmat})
would actually be a mistake. Indeed, in this case, the Fisher matrix
would include the derivatives of the lensed power spectra with respect
to the cosmological parameter.  So, the physical effect of each
cosmological parameter on lensing distortions would be counted several
times, not only in $\partial C_l^{dd} / \partial \theta_i$ but also in
$\partial C_l^{ab} / \partial \theta_i$, with $a,b \in \{ T, E \}$.
This would introduce correlations which would not be taken into
account self-consistently, and the forecasted errors would be
artificially small, as noticed in \cite{Kaplinghat:2003bh}.}.

Second, we assumed a perfect cleaning of all the astrophysical
foregrounds which contribute to the raw CMB observations. It is true
that CMB experiments are operating in various frequency bands,
precisely in order to subtract the foregrounds which frequency
dependence is usually non-planckian.  However, we still have a poor
knowledge of many foregrounds, and some of them could reveal very
difficult to remove, introducing extra non-gaussianity and spoiling
the lensing extraction process \cite{Amb04,Cooray:2005hm}.  In
particular, the question of foreground subtraction is related to the
maximum $l$ at which we should stop the sum in the Fisher matrix
expression, i.e.\ to the smallest angular scale on which we expect to
measure primary temperature and polarization anisotropies.  If we
assume a perfect cleaning, this value should be deduced from
instrumental noise. Beyond some multipoles ($l_{\rm max}^T$, $l_{\rm
max}^E$), the noise terms ($N_l^{TT}$, $N_l^{EE}$) become
exponentially large. Thus, in practice, the sum in Eq.~(\ref{trace})
can be stopped at any $l$ larger than both $l_{\rm max}^T$ and $l_{\rm
max}^E$. However, some foregrounds are expected to be impossible to
subtract on very small angular scales (e.g., radio sources, dusty
galaxies, or polarized synchrotron radiation and dust emission), so
for experiments with a very small instrumental noise, the covariance
matrix could be dominated by foreground residuals at smaller $l$
values than those where the instrumental noise explodes.

Since we do not have precise enough data at high galactic latitude and
on relevant frequencies, it is difficult at the moment to estimate how
problematic foreground contamination will be, but it is clear that one
should adopt a very careful attitude when quoting forecasted errors
for future experiments with an excellent angular resolution. In the
next sections, for each experiment and model, we will derive two
results: one optimistic forecast, assuming perfect foreground cleaning
up to the scale where the instrumental noise explodes (or in the case
of the most precise experiments, up to the limit $l_{\rm max}^T,
l_{\rm max}^E < 2750$ beyond which it is obvious that foreground
contamination will dominate); and one very conservative forecast,
assuming no foreground cleaning at all. In that case, we take the
foreground spectra $F_l^{TT}$, $F_l^{EE}$ and $F_l^{TE}$ of the
``mid-model'' of Ref.\ \cite{Tegmark:1999ke}, computed with the public
code provided by the authors\footnote{For each experiment, we compute
the foreground for each frequency channel, and then compute the
minimum variance combination of all components.}. This model is not
completely up-to-date, since it is based on the best data available at
the time of publication, and does not include important updates like
the level of polarized galactic dust observed by Archeops
on large angular scales \cite{Ponthieu:2005kv}. Also,
for simplicity, it assumes statistically
isotropic and Gaussian foregrounds, with no $TB$ or $EB$
correlations. 
However this approach is expected to provide the correct
orders of magnitude, which is sufficient for our purpose.  We add
these new terms to the covariance matrix of Eq.\ (\ref{lensed_covmat}),
as if they were extra noise power spectra for the ${TT}$, ${EE}$ and
${TE}$ pairs. We consistently recompute $N_l^{dd}$, still using the
equations in Ref.\ \cite{QE3} but with these extra noise terms
included, in order to model the worse possible loss of precision
induce by foregrounds in the lensing extraction process.  We expect
that the true error-bar for each cosmological parameter will be
somewhere between our two optimistic and conservative forecasts.

\section{Experimental sensitivities}

We consider seven CMB experiments which are representative of the
experimental efforts scheduled for the next decade.
The first two, based in the South Pole, are complementary:
BICEP\footnote{http://www.astro.caltech.edu/$\sim
$lgg/bicep$\_$front.html} (Background Imaging of Cosmic Extragalactic
Polarization) \cite{Keating} is designed for large angular scales,
while
QUaD\footnote{http://www.astro.cf.ac.uk/groups/instrumentation/projects/quad/}
(QUest at DASI, the Degree Angular Scale Interferometer) \cite{Church}
for small angular scales. The second experiment, which is already
collecting data, is composed of the QUEST (Q and U Extragalactic
Sub-mm Telescope) instrument mounted on the structure of the DASI
experiment.  A second set of experiments is scheduled in Antarctica at
the French-italian Concordia station and in the Atacama plateau in
Chile: the
BRAIN\footnote{http://apc-p7.org/APC$\_$CS/Experiences/Brain/index.phtml}
(B-modes Radiation measurement from Antarctica with a bolometric
INterferometer) \cite{Piat} instrument for measuring large scales, and
the ClOVER\footnote{http://www-astro.physics.ox.ac.uk/$\sim
$act/clover.html} (Cl ObserVER) \cite{Maffei} instrument for
intermediate scales.  BRAIN and ClOVER are designed for unprecedented
precision measurements of the $B$-mode for $l<1000$.  ClOVER was
approved for funding by PPARC in late 2004 and could be operational by
2008.  At that time, the {\sc Planck}\footnote{
http://sci.esa.int/science-e/www/area/index.cfm?fareaid=17 and
http://www.planck.fr/} satellite \cite{Tauber} should be collecting
data: {\sc Planck} has already been built and should be launched in
2007 by the European Space Agency (ESA). Beyond {\sc Planck}, at least
two space projects are under investigation: the mini-satellite SAMPAN
(SAtellite to Measure the Polarized ANisotropies) \cite{Bouchet}
for CNES (Centre National d'Etudes Spatiales), and the more ambitious
Inflation Probe project for NASA (National Aeronautics and Space
Administration), whose characteristics are not yet settled.  The
calculations of Ref.\ \cite{Kaplinghat:2003bh} were based on numbers
inspired from preliminary studies for the CMBpol satellite project: a
resolution of 1 arc-minute and a sensitivity per pixel of 1~$\mu$K for
temperature, 1.4 $\mu$K for polarization.  Here, the experiment that
we will call Inflation Probe is based on one over many possibilities
\cite{Bock}: a bolometer array with a passively cooled 3-4 m aperture
telescope, with four years of multifrequency observations and a
sensitivity of 2 $\mu$K s$^{-1/2}$ per channel.

\begin{table}
\begin{ruledtabular}
\begin{tabular}{cccccc}
Experiment & $f_{\rm sky}$ & $\nu$ & $\theta_b$ & $\Delta_T$ & $\Delta_E$\\
\hline
\hline
BICEP \cite{Keating} & 0.03
    & 100 & 60' & 0.33 & 0.47\\
&   & 150 & 42' & 0.35 & 0.49\\
QUaD \cite{Bowden:2003ub} & 0.025
    & 100 & 6.3' & 3.5 & 5.0\\
&   & 150 & 4.2' & 4.6 & 6.6\\
\hline
BRAIN \cite{Piat} & 0.03 
    & 100 & 50' & 0.23 & 0.33\\
&   & 150 & 50' & 0.27 & 0.38\\
&   & 220 & 50' & 0.40 & 0.56\\
ClOVER \cite{Maffei} & 0.018
    & 100 & 15' & 0.19 & 0.30\\
&   & 143 & 15' & 0.25 & 0.35\\
&   & 217 & 15' & 0.55 & 0.76\\
\hline
{\sc Planck} \cite{planck} & 0.65
    &  30 & 33'  &  4.4 &  6.2\\
&   &  44 & 23'  &  6.5 &  9.2\\
&   &  70 & 14'  &  9.8 & 13.9\\
&   & 100 & 9.5' &  6.8 & 10.9\\
&   & 143 & 7.1' &  6.0 & 11.4\\
&   & 217 & 5.0' & 13.1 & 26.7\\
&   & 353 & 5.0' & 40.1 & 81.2\\
%&   & 545 & 5.0' & 401  & $10^{10}$\\
&   & 545 & 5.0' & 401  & $\infty$\\
%&   & 857 & 5.0' & 18300 & $10^{10}$\\
&   & 857 & 5.0' & 18300 & $\infty$\\
\hline
SAMPAN \cite{Bouchet} & 0.65
  & 100 & 42' & 0.13 & 0.18\\
& & 143 & 30' & 0.16 & 0.22\\
& & 217 & 20' & 0.26 & 0.37\\
\hline
Inflation Probe       & 0.65 &  70 & 6.0' & 0.29 & 0.41 \\
{\em (hypothetical)} \cite{Bock}  
                      &      & 100 & 4.2' & 0.42 & 0.59 \\
                      &      & 150 & 2.8' & 0.63 & 0.88 \\
                      &      & 220 & 1.9' & 0.92 & 1.30 \\
\end{tabular}
\end{ruledtabular}
\caption{\label{tableexp}
Sensitivity parameters of the CMB projects considered in this work:
$f_{\rm sky}$ is the observed fraction of the sky,
$\nu$ the center frequency of the channels in GHz,
$\theta_b$ the FWHM (Full-Width at Half-Maximum) in arc-minutes, 
$\Delta_{T}$ the temperature
sensitivity per pixel in $\mu$K and $\Delta_E=\Delta_B$ the
polarization sensitivity. For all experiments, we
assumed one year of observations, except for
the Inflation Probe sensitivity based on four years.}
\end{table}

\begin{figure*}[ht]
\includegraphics[width=.45\textwidth]{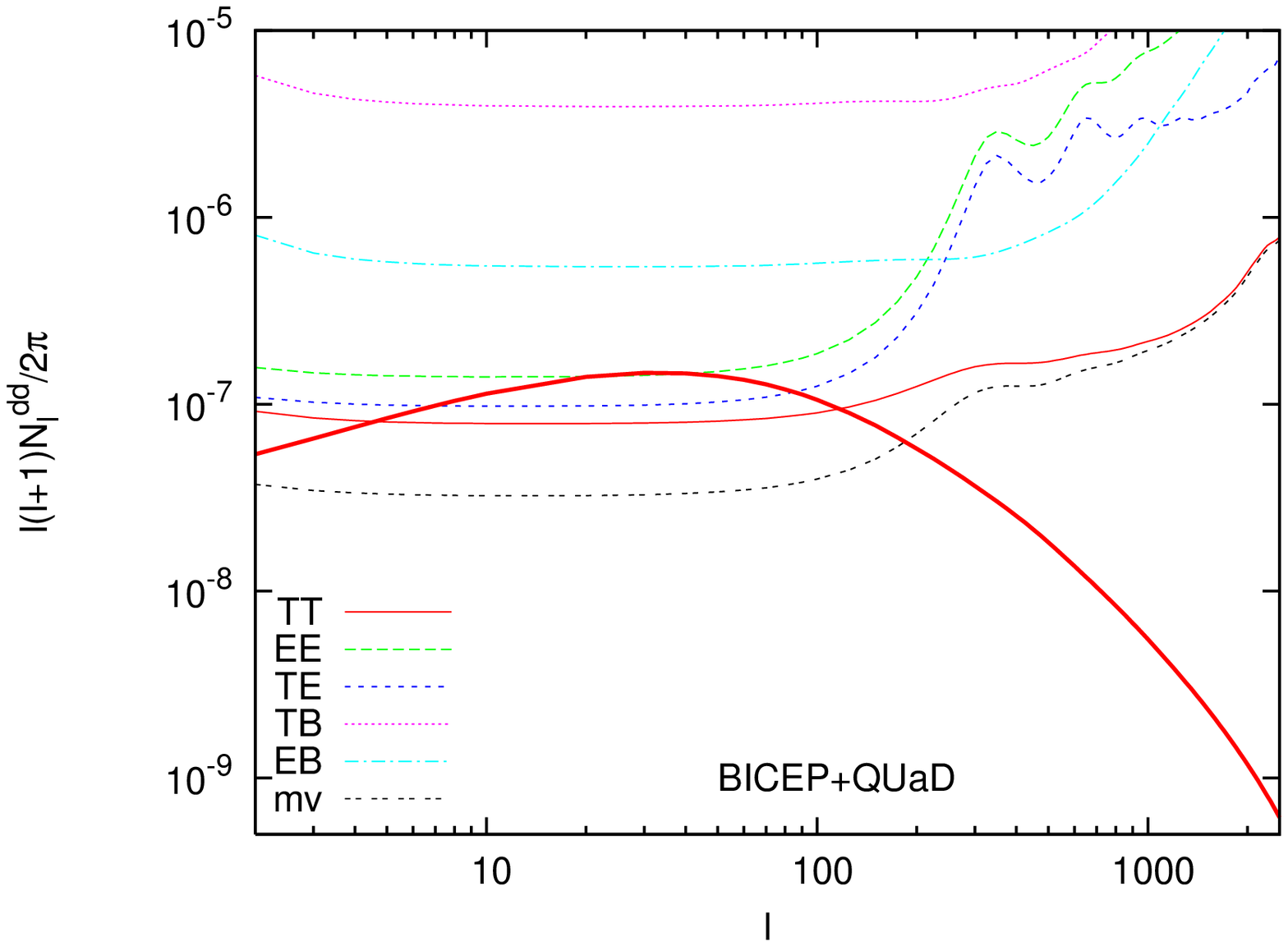}
\includegraphics[width=.45\textwidth]{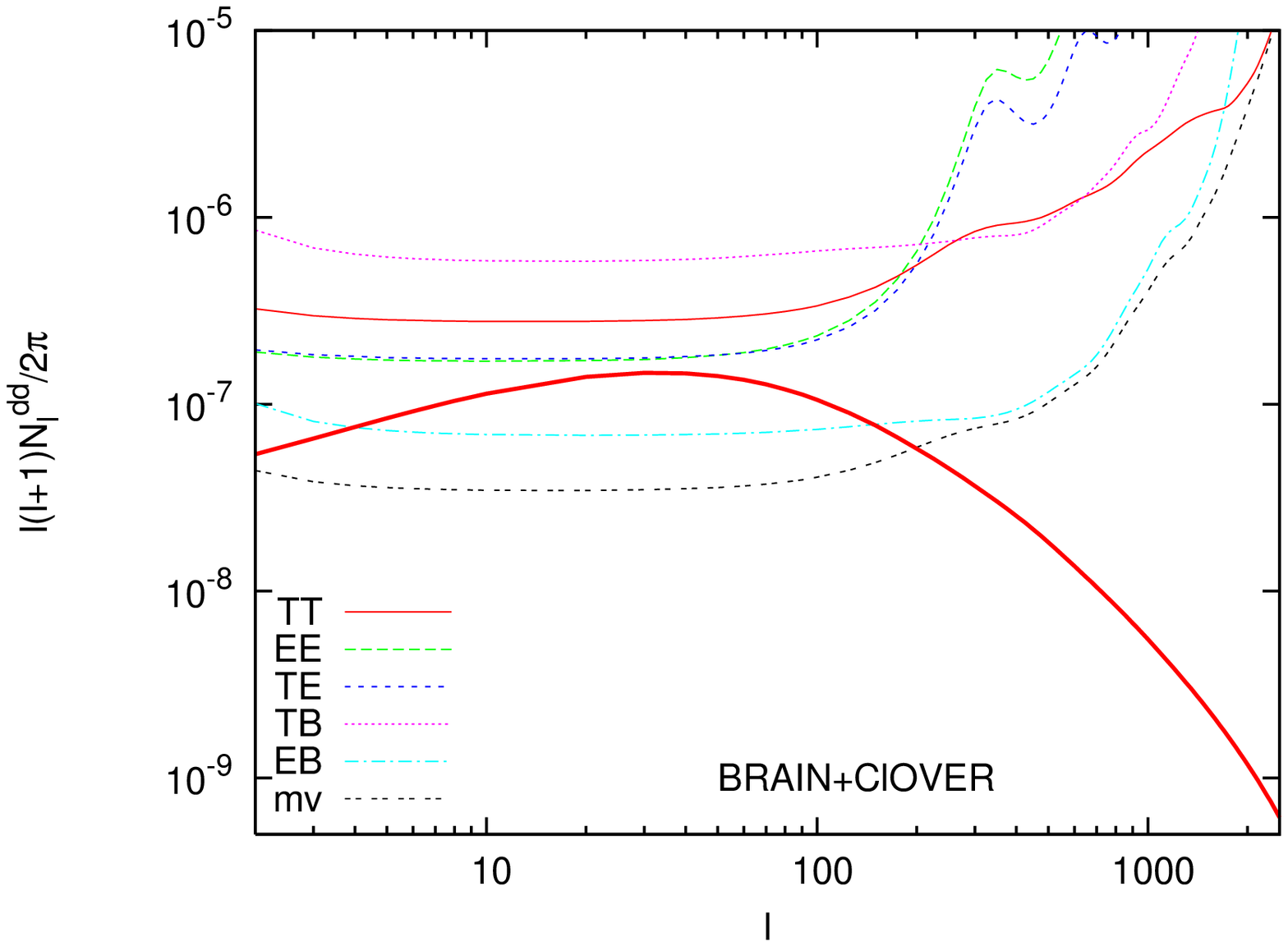}\\
\includegraphics[width=.45\textwidth]{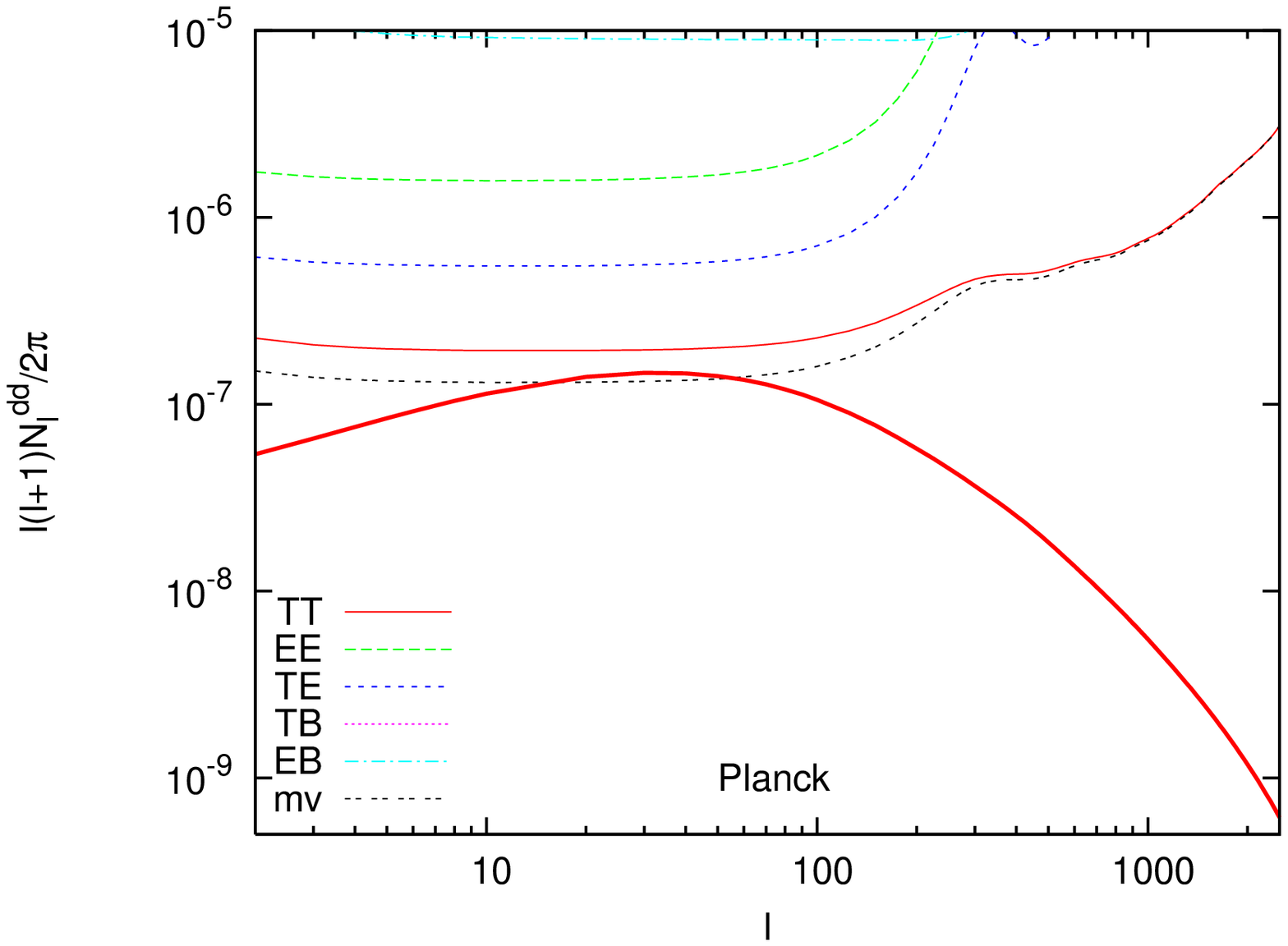}
\includegraphics[width=.45\textwidth]{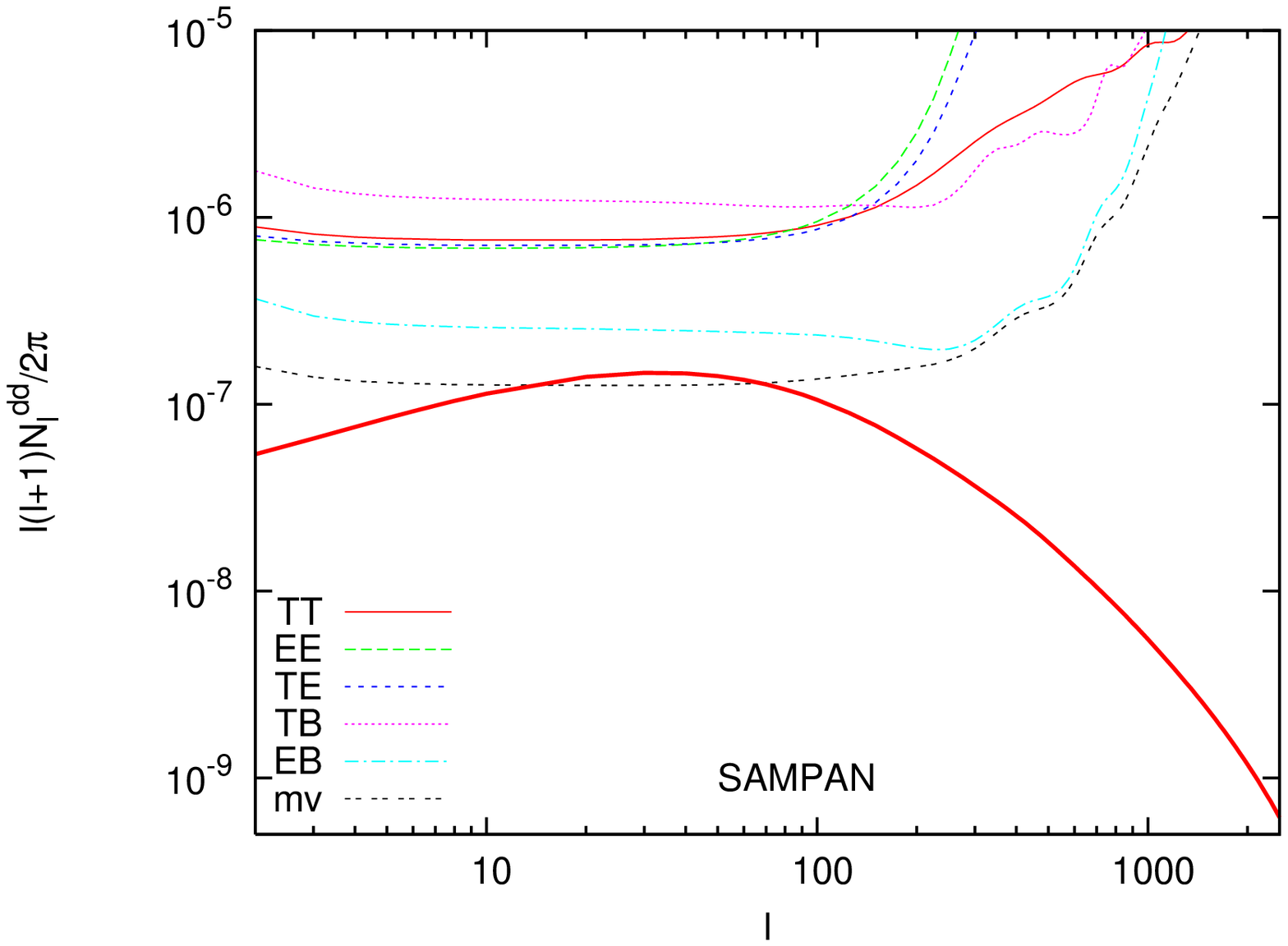}\\
\includegraphics[width=.45\textwidth]{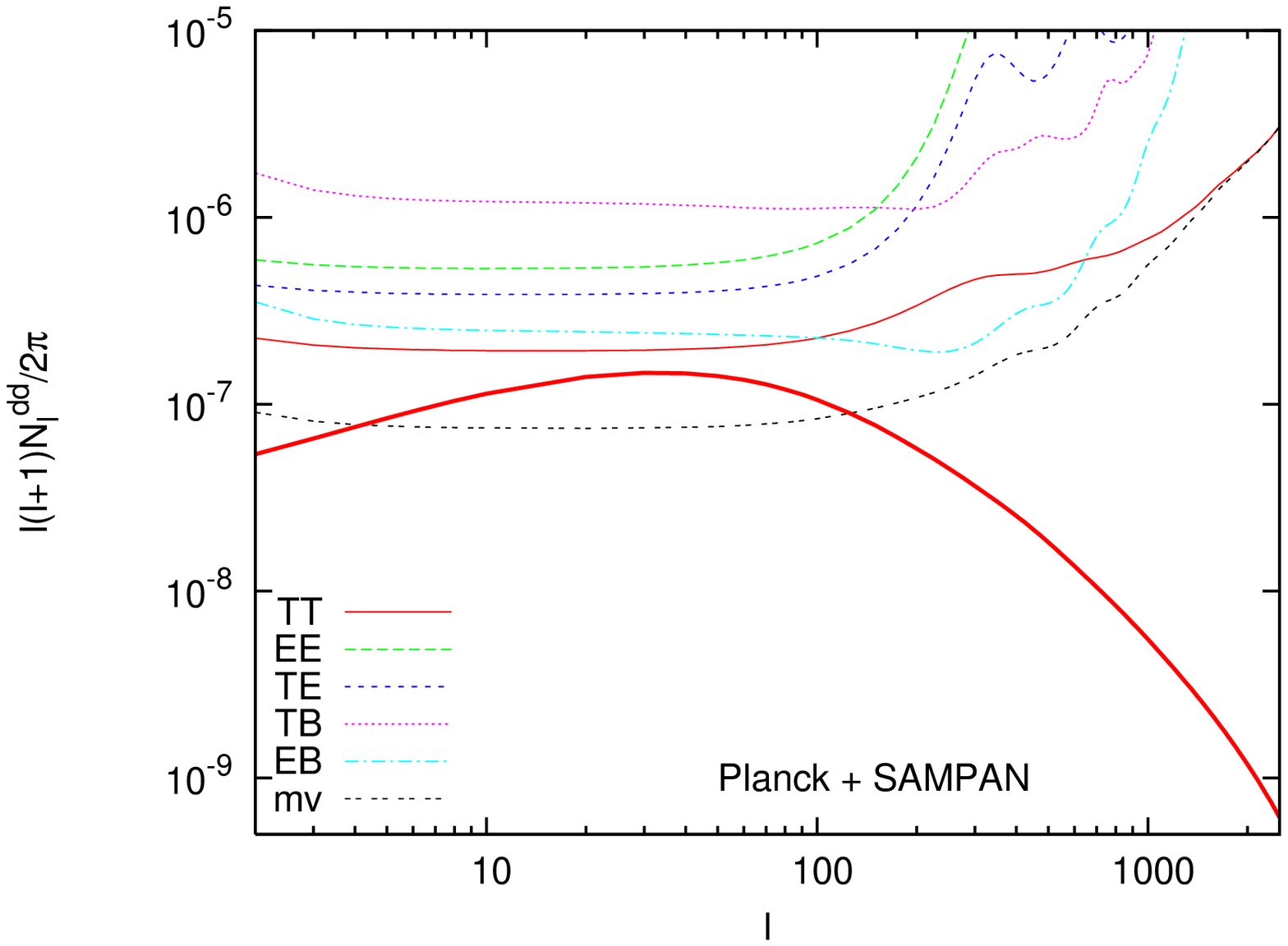}
\includegraphics[width=.45\textwidth]{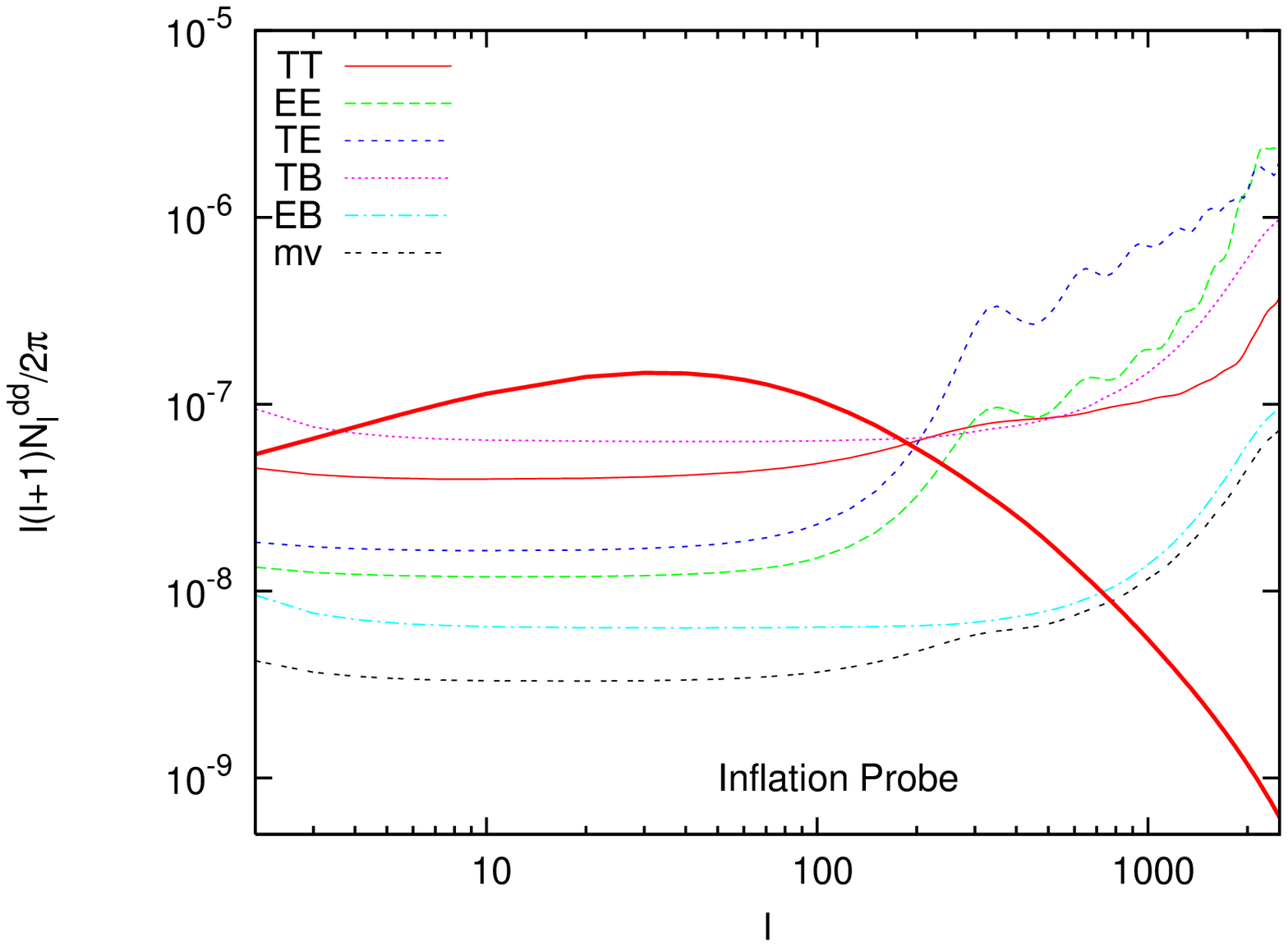}
\caption{\label{fig_dd_errors} For six CMB experiments or
combinations of experiments, we show the expected noise power
spectrum $N_l^{dd}$ for the quadratic estimators $d(a,b)$ built out of
pairs $ab \in \{TT, EE, TE, TB, EB\}$, and for the combined minimum
variance estimator (mv). The thick line shows for comparison 
the signal power spectrum 
$C_l^{dd}=\langle d_l^m d_l^{m*} \rangle$. The sum of the two curves
$N_l^{dd}+C_l^{dd}$ represents the expected variance 
of a single multipole $d(a,b)_l^m$.}
\end{figure*}

\begin{figure*}[ht]
\includegraphics[width=.45\textwidth]{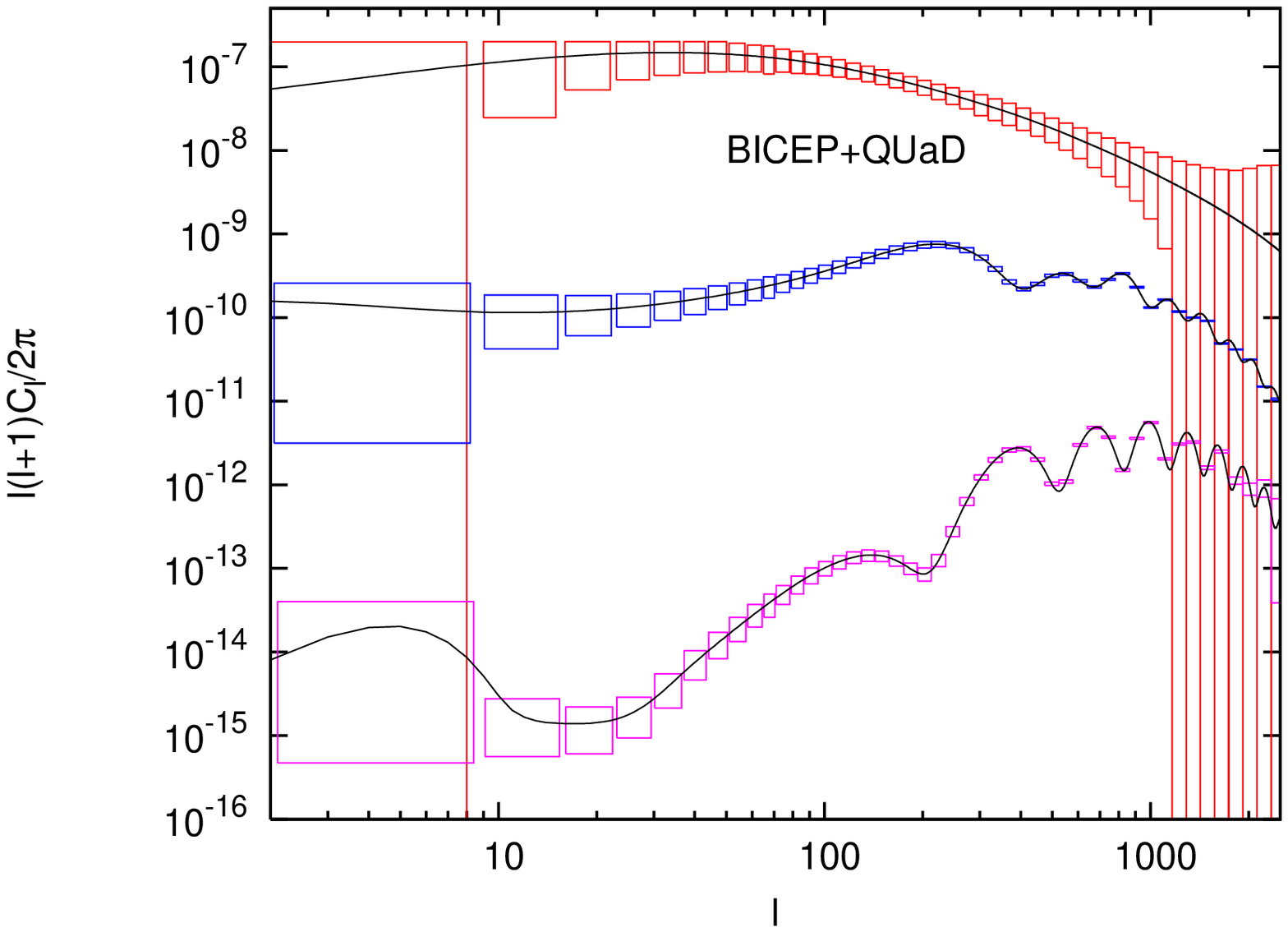}
\includegraphics[width=.45\textwidth]{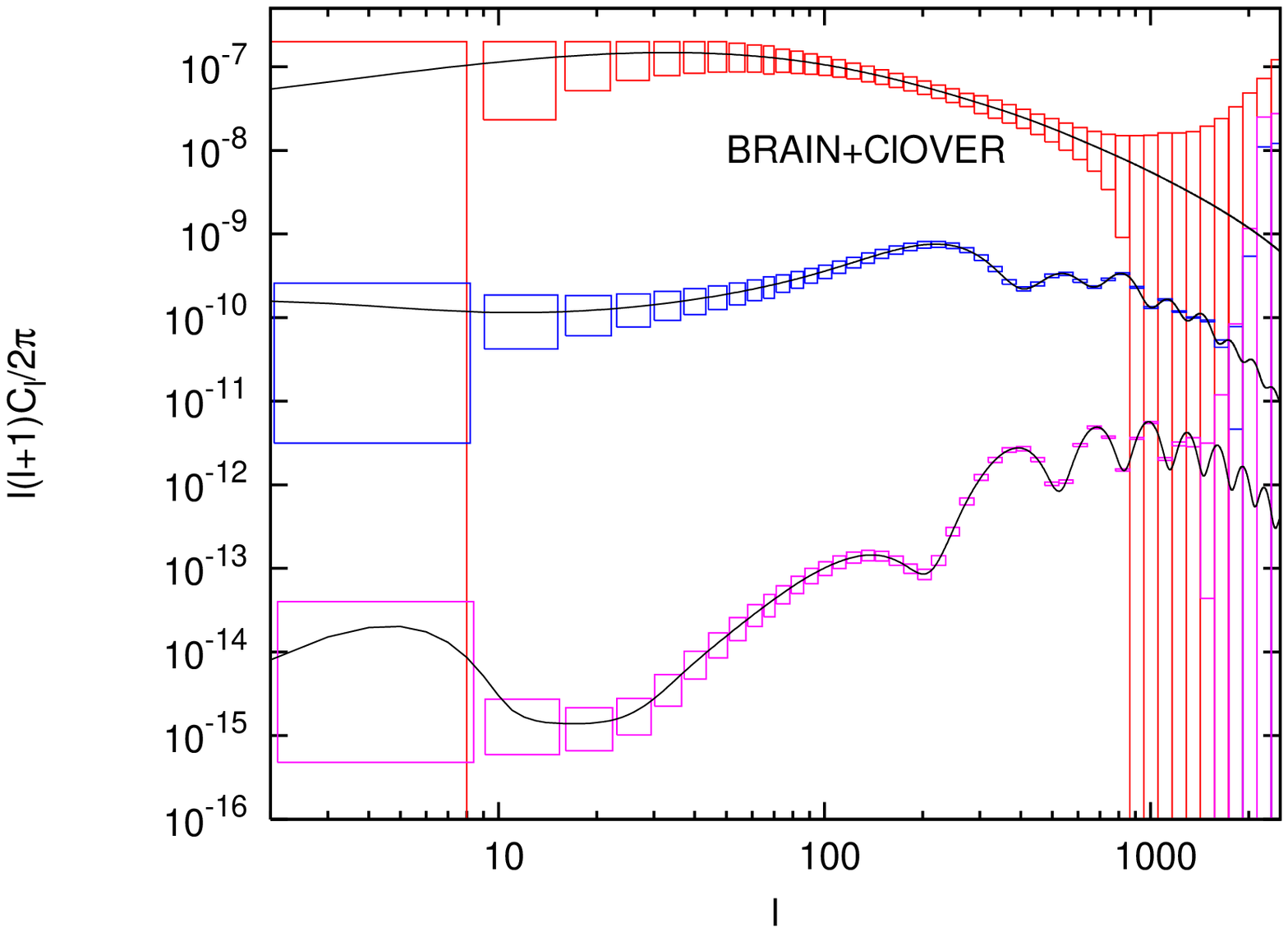}\\
\includegraphics[width=.45\textwidth]{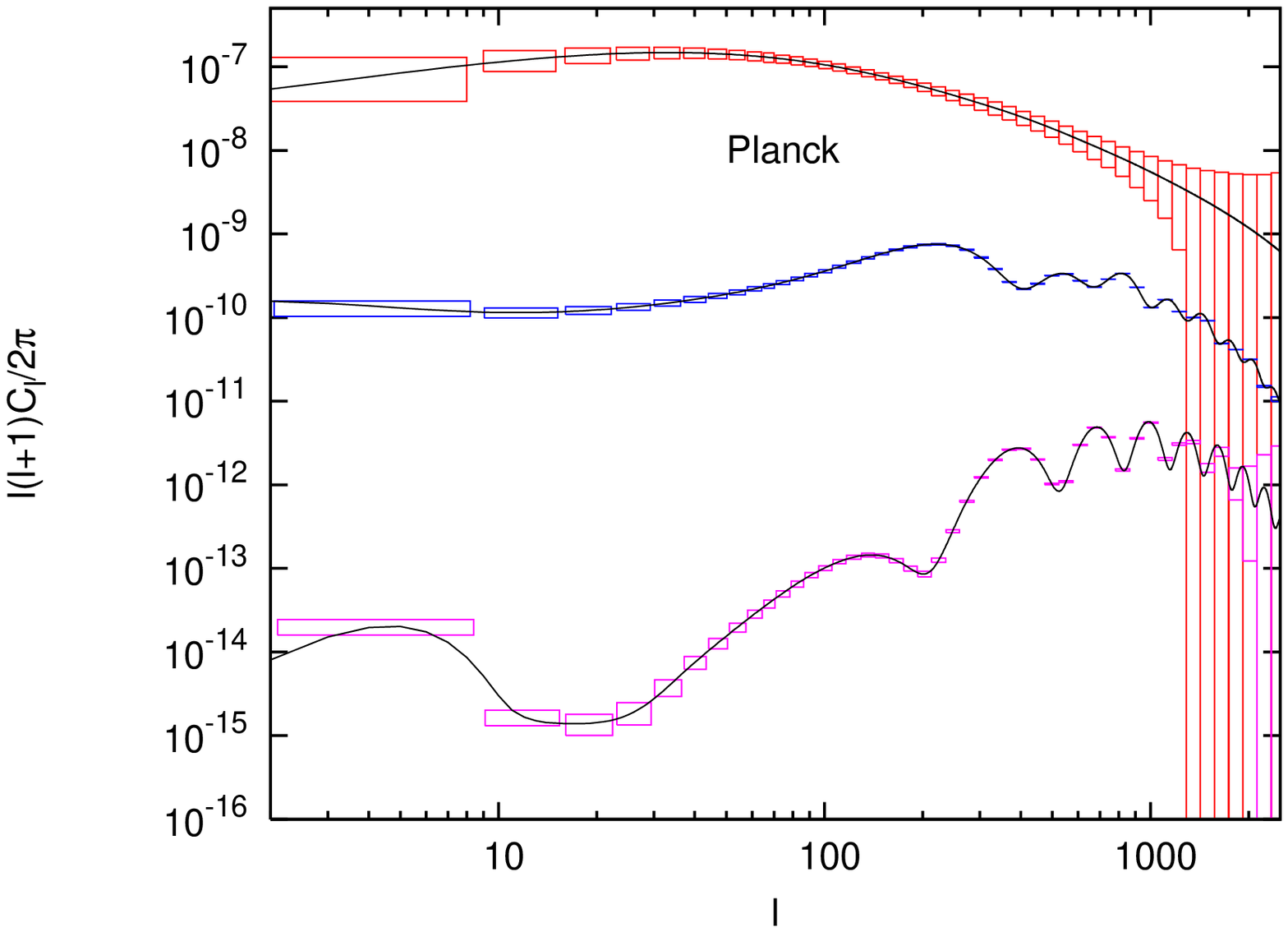}
\includegraphics[width=.45\textwidth]{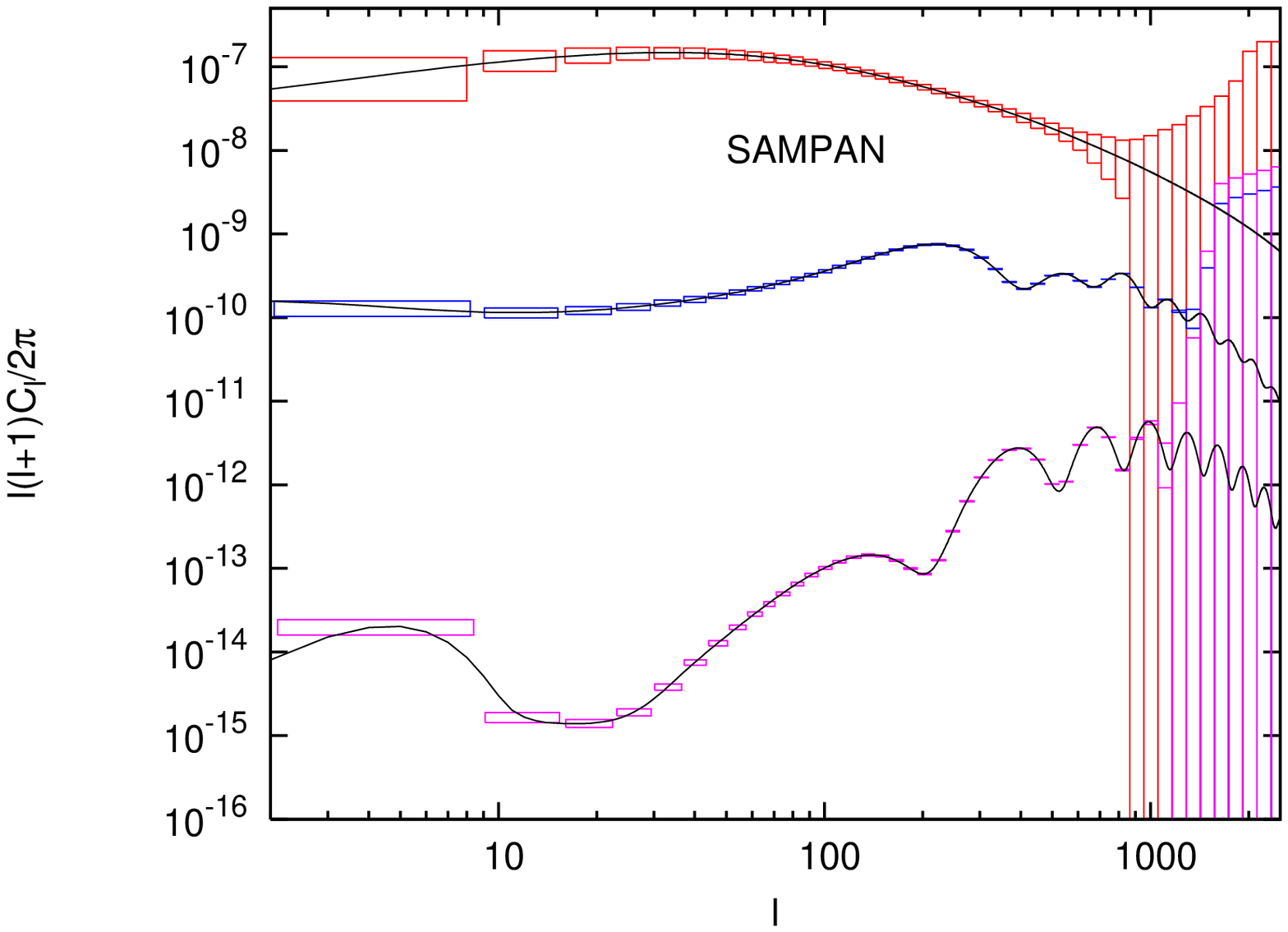}\\
\includegraphics[width=.45\textwidth]{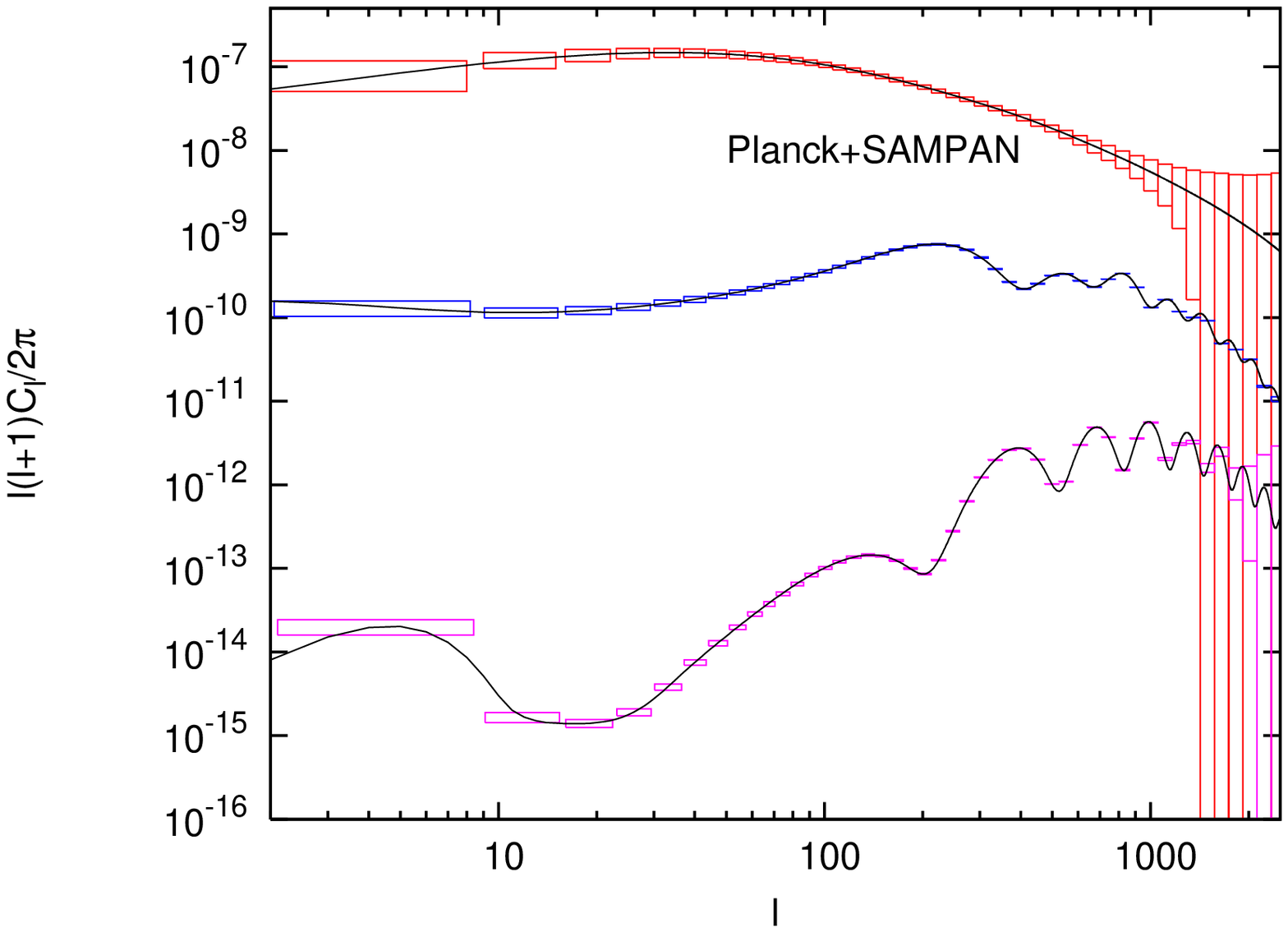}
\includegraphics[width=.45\textwidth]{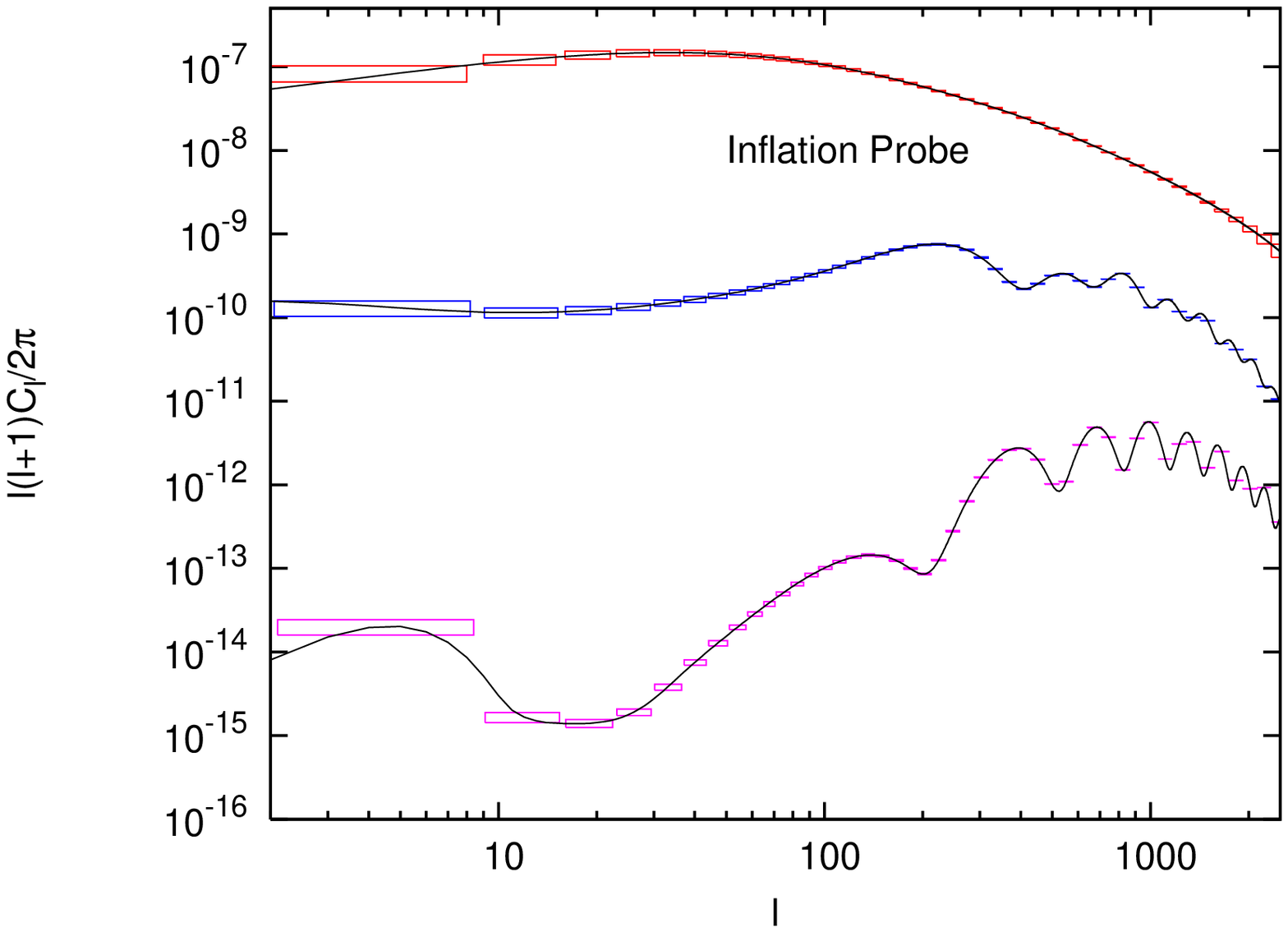}
\caption{\label{fig_binned_errors} For the same six CMB experiments or
combinations of experiments as in figure \ref{fig_binned_errors}, we
show the expected binned error on the reconstructed power spectra:
from top to bottom, $C_l^{dd}$ (using the minimum variance quadratic
estimator), $C_l^{TT}$ and $C_l^{EE}$. The curves represent the power
spectra of the fiducial model described in section \ref{sec:results}.}
\end{figure*}

We list the expected instrumental performances of each experiment in
Table \ref{tableexp}.  Each instrument includes many detectors grouped
in frequency bands or {\it channels}. In each channel, the detectors
have a given spatial resolution described by the FWHM (Full-Width at
Half-Maximum) $\theta_b$ of the beam. For a given channel, one can
estimate the temperature and polarization sensitivities per pixel
of the combined detectors, $\Delta_{T}$ and 
$\Delta_{E}=\Delta_{B}$.
The channel noise power spectrum reads
\begin{equation}
N_{l, \nu}^{aa}=(\theta_b \Delta_a)^2 
\exp \left[ l(l+1)\theta_b^2/8\ln 2 \right]~,
\end{equation}
with $a \in \{ T, E, B \}$.  The noise from individual channels can be
combined into the global noise of the experiment
\begin{equation}
N_l^{aa}=\left[ \sum_{\nu} (N_{l, \nu}^{aa})^{-1} \right]^{-1}.
\end{equation}
Given this input, the computation of the lensing noise $N_l^{dd}$ can
be performed numerically following Ref.~\cite{QE3}.  In
Fig.~\ref{fig_dd_errors}, we show our results for the lensing noise of
each experiment, based on each quadratic estimator and on the combined
minimum variance estimator. In Fig.~\ref{fig_binned_errors} we gather
information on the noise for the $TT$, $EE$ and $dd$ power spectra for
each experiment.  The error-bars $\Delta C_l^{aa}$ displayed in
Fig.~\ref{fig_binned_errors} include both cosmic variance and
instrumental noise, and assume a multipole binning of width $\Delta l
= 7$ until $l \sim 70$, and then $\Delta l \sim l/10$
\begin{equation}
\Delta C_l^{aa} = \sqrt{\frac{2}{(2l+1) \, \Delta l \, f_{\rm sky}}}
(C_l^{aa} + N_l^{aa})~.
\end{equation}
The top graphs in Figs.\ \ref{fig_dd_errors} \&
\ref{fig_binned_errors} correspond to the BICEP+QUaD and BRAIN+ClOVER
combinations.  Computing the Fisher matrix for each pair of
experiments is not a trivial task, due to the different sky
coverages. We follow a method which is certainly not optimal, but has
the merit of simplicity.  Since in each case, one experiment is
optimized for large scales and the other for smaller scales, we assume
that below a given value $l_c$ all multipoles are evaluated from BICEP
or BRAIN only, while for $l>l_c$ they are taken from QUaD or
ClOVER. In Eqs.\ (\ref{trace}, \ref{lensed_covmat}), this amounts in
considering $f_{\rm sky}$ as a function of $l$, and in replacing
$f_{\rm sky}(l)$, $N_l^{TT}$ and $N_l^{EE}$ by their BICEP/BRAIN value
for $l<l_c$, or by their QUaD/ClOVER value for $l>l_c$. The lensing
noise $N_l^{dd}$ is then computed for the combined experiment,
following the same prescriptions.  For each pair of experiments, we
optimized the value of $l_c$ numerically by minimizing the forecasted
error on the total neutrino mass $M_{\nu}$.  In both cases, we found
that $l \sim 300$ was optimal. This method might be less favorable for
BRAIN+ClOVER than for BICEP+QUaD, because the first pair of
experiments has a large overlap in $l$-space, for which multipoles
could be derived from the two combined datasets.

We find that BICEP+QUaD is able to reconstruct the lensing multipoles
$d_l^m$ in the range $2 < l < 200$ with an impressively small noise
power spectrum $N_l^{dd}$. QUaD has both an excellent resolution and a
very good sensitivity, and should provide an extremely precise
measurement of $T$ and $E$ modes on small angular scales. Therefore,
the three quadratic estimators $d(T,T)$, $d(E,E)$ and $d(T,E)$ are
particularly efficient.

The main goal of the BRAIN+ClOVER combined experiment is to improve
the determination of the $B$-mode performed by BICEP+QUaD, especially
on large and intermediate scales ($l < 1000$), which are particularly
important for detecting gravity waves from inflation.  This should be
achieved with a sensitivity which is even better than that of BICEP
and QUaD, but at the expense of a poorer resolution in the case of
ClOVER, leading to large errors for small-scale polarization. In
total, this design is roughly equivalent to that of BICEP+QUaD in
terms of lensing extraction: BRAIN+ClOVER is also able to reconstruct
the lensing multipoles $d_l^m$ in the range $2<l< 200$. The best
estimator is now $d(E,B)$, known to be particularly useful, since $E$
and $B$ are correlated only due to lensing. In this sense, future
lensing determinations by BRAIN+ClOVER and by BICEP+QUaD can be seen
as complementary, and therefore both particularly interesting.

The {\sc Planck} satellite has a resolution comparable to QUaD, but a
poorer sensitivity than the last four experiments. This explains why
the lensing noise shown in Fig.\ \ref{fig_dd_errors} looks a bit
disappointing: the signal marginally exceeds the noise only around $l
\simeq 40$.  However, we should keep in mind that {\sc Planck} will
observe the full sky (which leads to $f_{\rm sky}=0.65$, once the 
galactic cut has been taken into account), while BICEP+QUaD or
BRAIN+ClOVER explore only small regions. Therefore, for a given $l$,
{\sc Planck} makes many more independent measurements of multipoles
($T_l^m$, $E_l^m$), and consequently, also of $d_l^m$. In Fig.\
\ref{fig_binned_errors}, one can check that {\sc Planck} still makes a
more precise determination of the lensing power spectrum than
BICEP+QUaD: both experiments are able to constrain $C_l^{dd}$ up to $l
\sim 1100$, but the satellite provides smaller errors.

Since {\sc Planck} is not very sensitive to $B$-modes, and BRAIN is
limited by its small sky coverage, there will be room after these two
projects for improving $B$-mode observations on large angular scales,
in view of observing inflationary gravitational waves.  This would be
the target of the SAMPAN mini-satellite project, which would be a
full-sky experiment with excellent sensitivity but poor resolution.
We find that for the minimum variance estimator, the noise $N_l^{dd}$
would be at the same level for {\sc Planck} and SAMPAN. However, it is
interesting to note that Sampan has a good $d(E,B)$ estimator, while
{\sc Planck} is better with $d(T,T)$. Therefore, it sounds
particularly appealing to combine the two full-sky experiments, that
is technically equivalent to assuming a super-experiment with twelve
channels (nine from {\sc Planck} and three from SAMPAN).  The results
(in the fifth graphs of Figs.~\ref{fig_dd_errors} \&
\ref{fig_binned_errors}) show that with such a combination one could
lower the noise $N_l^{dd}$ by a factor two for the minimum variance
estimator, in order to constrain $C_l^{dd}$ up to $l \sim 1300$.

Finally, the (hypothetical) version of the Inflation Probe satellite
that we consider here has an extremely ambitious resolution and
sensitivity, such that the instrumental error would be better than
cosmic variance for the $B$-mode until $l \sim 1500$.  For such a
precise experiment, assumptions concerning foreground subtraction play
a crucial role, since it is very likely that foreground residuals will
start dominating the observed power spectrum before instrumental
noise. The last graphs in Figs.~\ref{fig_dd_errors} \&
\ref{fig_binned_errors}, which assume perfect foreground cleaning up
to $l \sim 2500$, show that lensing multipoles $d_l^m$ could be
recovered up to to $l\sim 800$, while $C_l^{dd}$ could be constrained
up to at least $l \sim 2500$.

\section{Future sensitivities to neutrino masses}
\label{sec:results}

\begin{table*}[ht]
\begin{ruledtabular}
\begin{tabular}{lcccccccc}
Free parameters:     
& \multicolumn{4}{c}{8 parameters of minimal $\Lambda$MDM} 
& \multicolumn{4}{c}{same + \{$\alpha$, $w$, $N_{\rm eff}$\} } \\
\hline
Lensing extraction:  & no & no & yes & yes & {no} & {no} & {yes} & {yes} \\
\hline
Foreground cleaning: & perfect & none & perfect & none & {perfect} & {none} & {perfect} & {none} \\
\hline
QUaD+BICEP   &1.3 &1.6 &0.31&0.36&{1.5 }&{1.9 }&{0.36}&{0.40}\\
BRAIN+ClOVER &1.5 &1.8 &0.34&0.43&{1.7 }&{2.0 }&{0.42}&{0.51}\\
{\sc Planck}       &0.45&0.49&0.13&0.14&{0.51}&{0.56}&{0.15}&{0.15}\\
SAMPAN       &0.34&0.40&0.10&0.17&{0.37}&{0.44}&{0.12}&{0.18}\\
{\sc Planck}+SAMPAN&0.32&0.36&0.08&0.10&{0.34}&{0.40}&{0.10}&{0.12}\\
Inflation Probe  &0.14&0.16&0.032&0.036&{0.25}&{0.26}&{0.035}&{0.039}\\
\end{tabular}
\end{ruledtabular}
\caption{\label{tableresults} Expected 1-$\sigma$ error on the total
neutrino mass $M_{\nu}$ in eV for various CMB experiments or
combinations of them. The first (last) four columns correspond to a
$\Lambda$MDM model with eight (eleven) free parameters. For each of
the two models, the four columns show the cases with or without
lensing extraction, and with two extreme assumptions concerning the
foreground treatment: perfect subtraction or no subtraction at all.}
\end{table*}

For each experiment, we compute the Fisher matrix following Eqs.\
(\ref{trace}, \ref{lensed_covmat}), for a $\Lambda$MDM fiducial model
with the parameter values as given below, and considering two
possibilities for the number of free parameters that should be
marginalized out.

The first possibility is the minimal alternative on the basis of
current observations: we marginalize over eight free parameters,
standing for the current baryon density $\omega_{\rm b}=\Omega_{\rm
b}\,h^2$, the current total matter density $\omega_{\rm m}=\Omega_{\rm
m}\,h^2$, the current dark energy density $\Omega_{\Lambda}$, the
total neutrino mass $M_{\nu}$ in eV, the primordial curvature power
spectrum amplitude $A_s$ and tilt $n_s$, the optical depth to
reionization $\tau$ and the primordial helium fraction $y_{\rm He}$,
to which we assign the values $(\omega_{\rm b}, \omega_{\rm m},
\Omega_{\Lambda}, M_{\nu}, \ln[10^{10} A_s], n_s, \tau, y_{\rm
He})=(0.023, 0.143, 0.70, 0.1, 3.2, 0.96, 0.11, 0.24)$. We assume no
spatial curvature and tensor contribution.  Note that the reduced
Hubble parameter derives from $h=\sqrt{\omega_{\rm
m}/(1-\Omega_{\Lambda})}$.

The second possibility, describing non-minimal physical assumptions,
is to marginalize over three extra parameters: the scalar tilt running
$\alpha$, which can be non-negligible in some inflationary models with
extreme assumptions; the dark energy equation-of-state parameter $w$;
and finally, extra relativistic degrees of freedom which would enhance
the total radiation density, parametrized by the effective number of
neutrino species $N_{\rm eff}$ (for instance, $N_{\rm eff}=4$ means
that the Universe contains a background of extra relativistic
particles with the same density as one extra massless neutrino
species). In the fiducial model, these parameters take the values
$(\alpha, w, N_{\rm eff})=(0, -1, 3)$. Our purpose is to find out
whether such extra free parameters open up degeneracy directions in
parameter space, that would worsen the sensitivity to neutrino masses.
It has been shown in recent analyses that these parameter degeneracies
indeed appear with current CMB and LSS data (see
\cite{Hannestad:2003ye,Crotty:2004gm} for $N_{\rm eff}$ and
\cite{Hannestad:2005gj,Ichikawa:2005hi} for $w$).

\begin{figure}%[ht]
\includegraphics[width=0.5\textwidth]{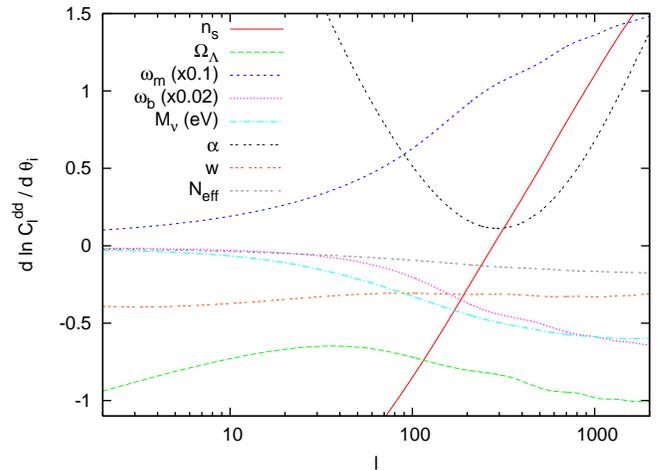}
\caption{\label{fig_deriv_dd} 
Logarithmic derivatives of the lensing power spectrum $C_l^{dd}$ with respect
to each cosmological parameter. The derivatives with respect
to $\omega_{\rm b}$ and $\omega_{\rm m}$ have been rescaled in order 
to fit inside the figure.}
\end{figure}

The derivative of the lensing power spectrum $C_l^{dd}$ with respect
to each of these eleven parameters are shown in Fig.\
\ref{fig_deriv_dd}, with the exception of the derivatives with respect
to $\tau$ and $y_{\rm He}$ which are null, and with respect to $A_s$
which is trivial. All derivatives were computed using the public
Boltzmann code {\sc camb} \cite{Lewis:1999bs}, enabling the highest
accuracy options and increasing the {\tt accuracy\_boost} parameter to
five. Whenever possible, we evaluated double-sided derivatives, and
searched for optimal step sizes such that the results were not
affected by numerical errors (from the limited precision of the code)
nor by contributions from higher-order derivatives.

We quote the results for the total neutrino mass $M_{\nu}$ in Table
\ref{tableresults}, assuming either eight or eleven free
parameters. For each of the two cases, we compare the forecasted
errors with and without lensing extraction, i.e.\ using either a
$2\times2$ or a $3\times3$ data covariance matrix, in order to
evaluate the impact of the extraction technique. Finally, in each of
the four sub-cases, we quote the results obtained assuming perfect
foreground cleaning or no cleaning at all, in order to be sure to
bracket the true error. Should we trust more the results based on the
eight or eleven parameter model?  This depends on future results from
cosmological observations: in absence of strong observational
motivation for extra parameters, one will probably prefer to stick to
the simplest paradigm; however, the next years might bring some
surprises, like for instance the detection of a variation in the dark
energy density.

Let us comment the results for each experiment. 
The combination QUaD+BICEP benefits a lot from lensing extraction,
since the error decreases from approximately $1.5$ eV to at least
$0.4$ eV.  These results are found to be robust against foreground
residuals and extra parameter degeneracies.  It is interesting that
with QUaD+BICEP it should soon be possible to reach in a near future
--using CMB only-- the same precision that we have today combining
many observations of different types (galaxy-galaxy correlation
function, Lyman-$\alpha$ forests) which are affected by various
systematics.  The situation is almost the same for BRAIN+ClOVER, which
should also achieve $\sigma(M_{\nu}) \sim 0.4$ eV using lensing
extraction.

{\sc Planck} should make a decisive improvement, lowering the error to
$\sigma(M_{\nu}) \sim 0.15$ eV, in excellent agreement with the
results of Ref.\ \cite{Kaplinghat:2003bh}. Note that without lensing
extraction the error would be multiplied by three (by four in the case
with extra free parameters). We do not find a significant difference
between the forecasted errors in the eight and eleven parameter
models. SAMPAN alone is slightly more efficient than {\sc Planck}, and
the combination {\sc Planck}+SAMPAN is the first one to reach
$\sigma(M_{\nu}) \sim 0.1-0.12$ eV, even in the pessimistic case of large
foreground residuals and extra free parameters.  Thus these future CMB
lensing data could help in breaking the parameter degeneracy between
$M_{\nu}$ and $w$ \cite{Hannestad:2005gj}, that would still be
problematic at the level of precision of Planck (without lensing
extraction) combined with the galaxy-galaxy correlation function
extracted from the Sloan Digital Sky Survey.

\begin{figure}%[ht]
\includegraphics[width=.23\textwidth]{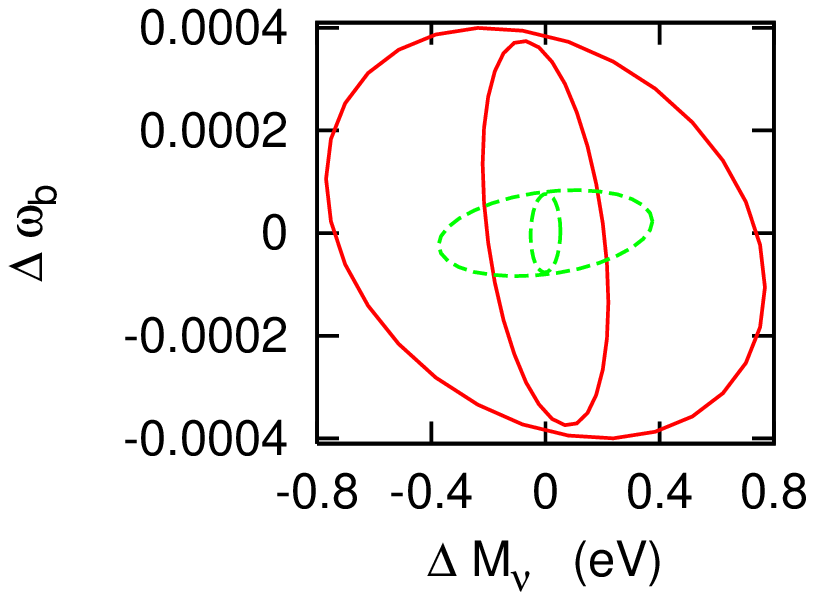}
\includegraphics[width=.23\textwidth]{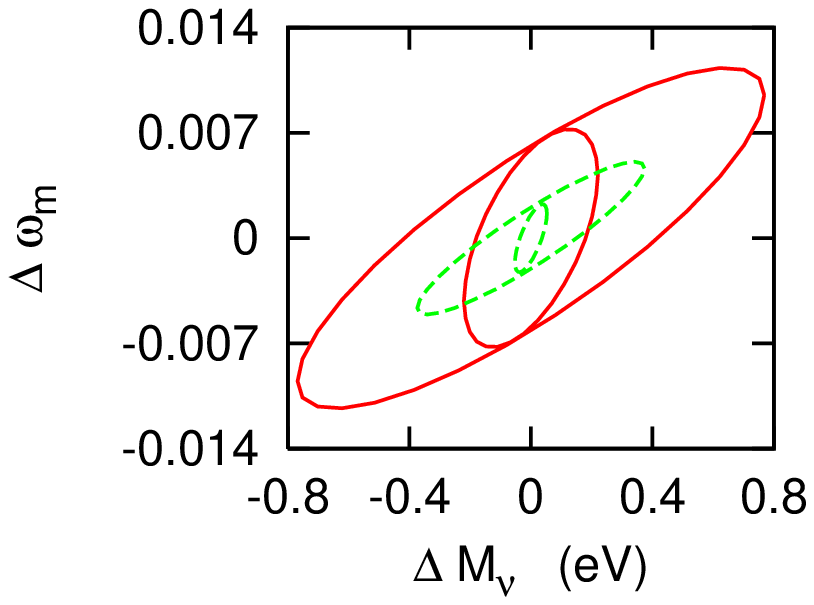}\\
\includegraphics[width=.23\textwidth]{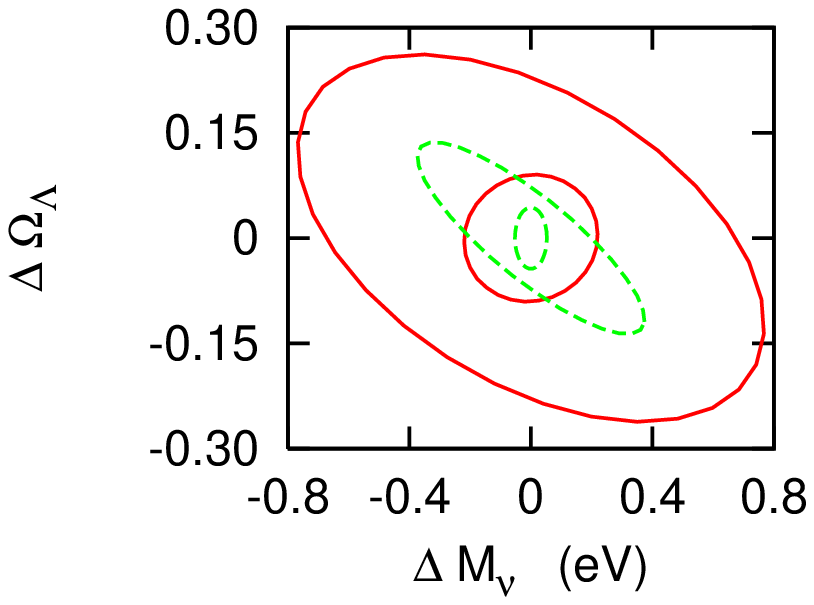}
\includegraphics[width=.23\textwidth]{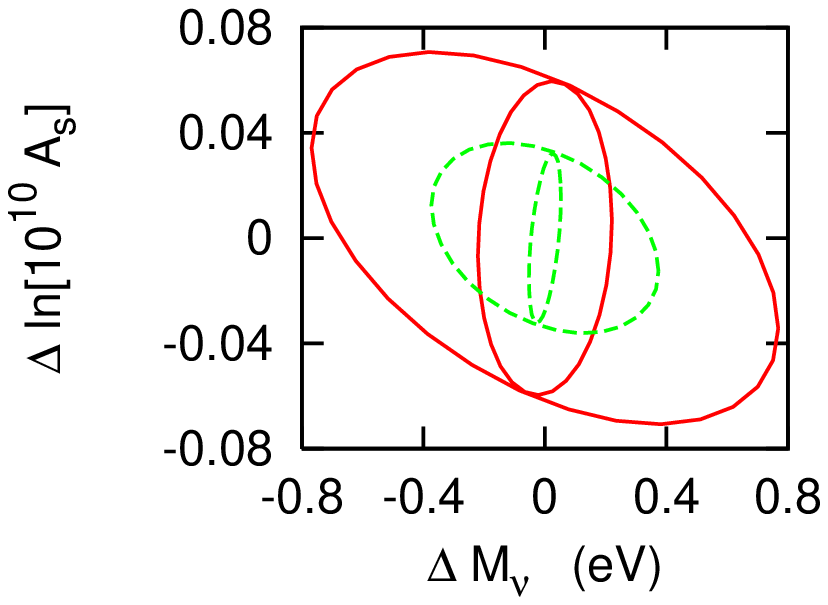}\\
\includegraphics[width=.23\textwidth]{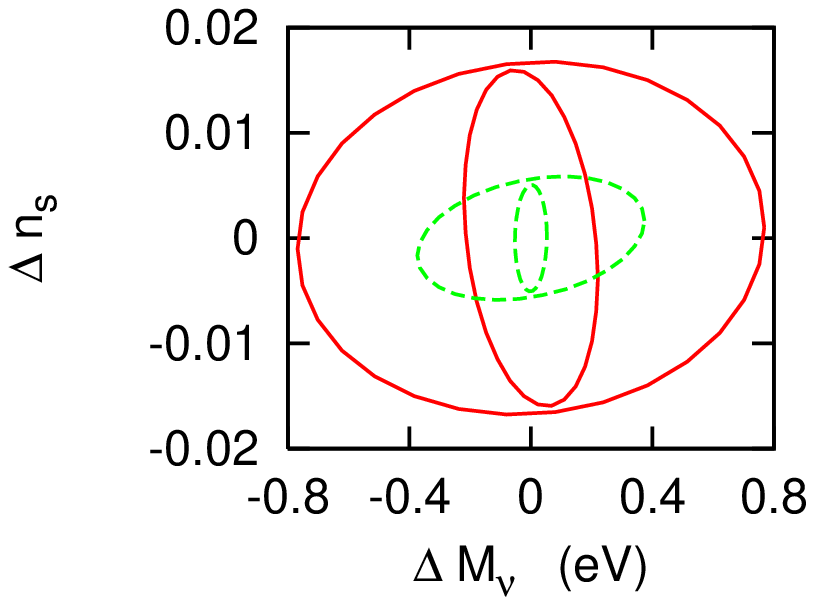}
\includegraphics[width=.23\textwidth]{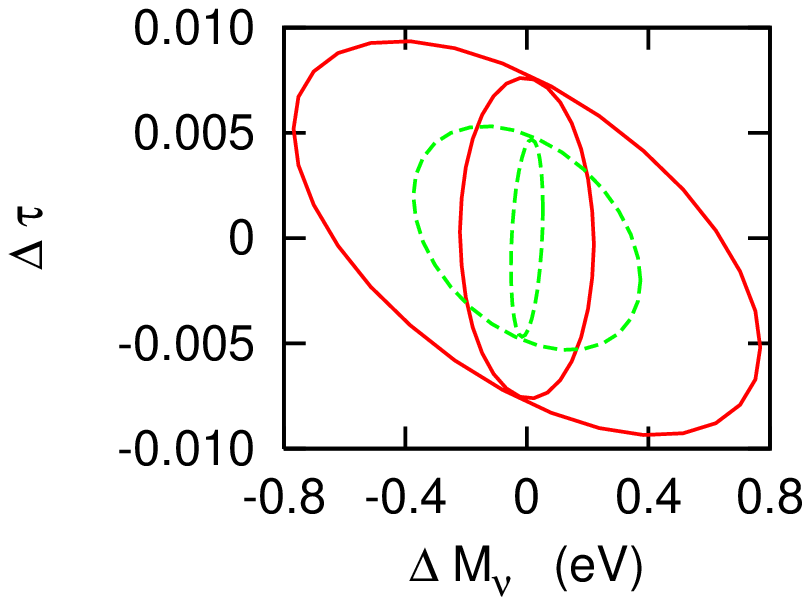}\\
\includegraphics[width=.23\textwidth]{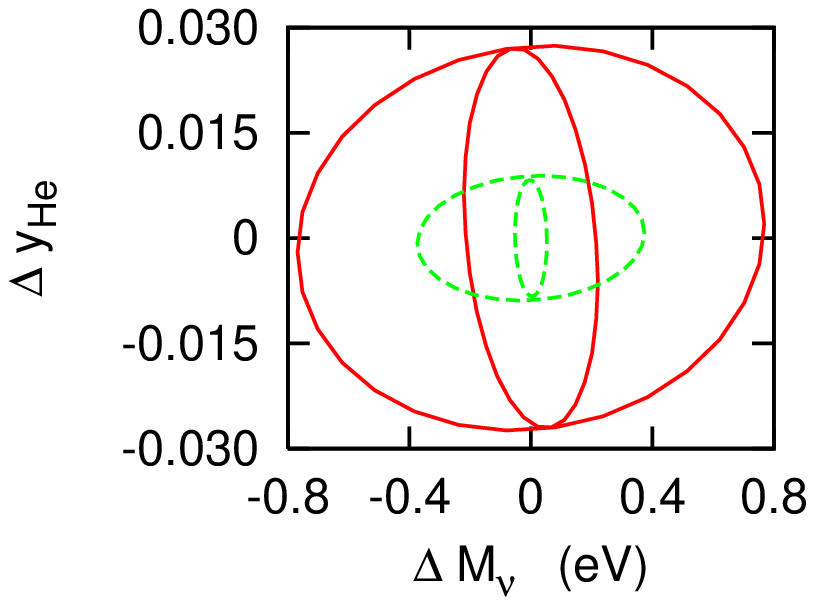}
\includegraphics[width=.23\textwidth]{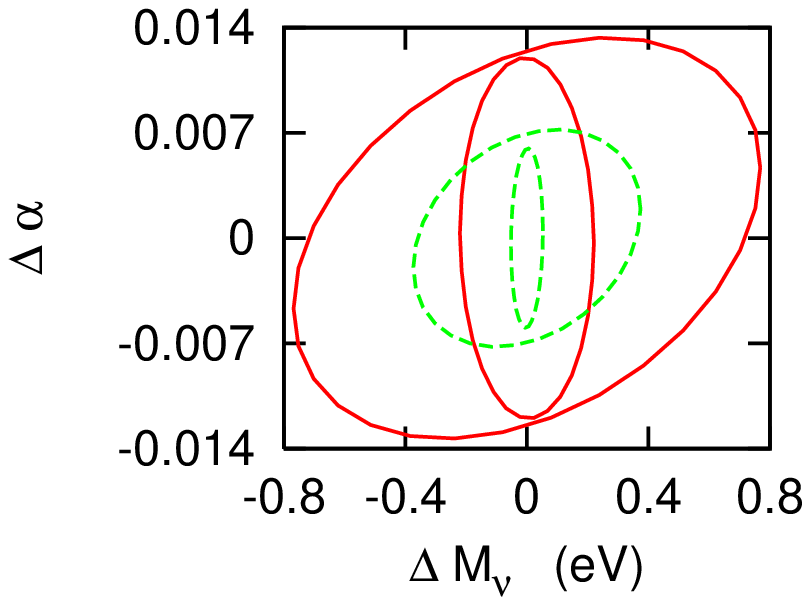}\\
\includegraphics[width=.23\textwidth]{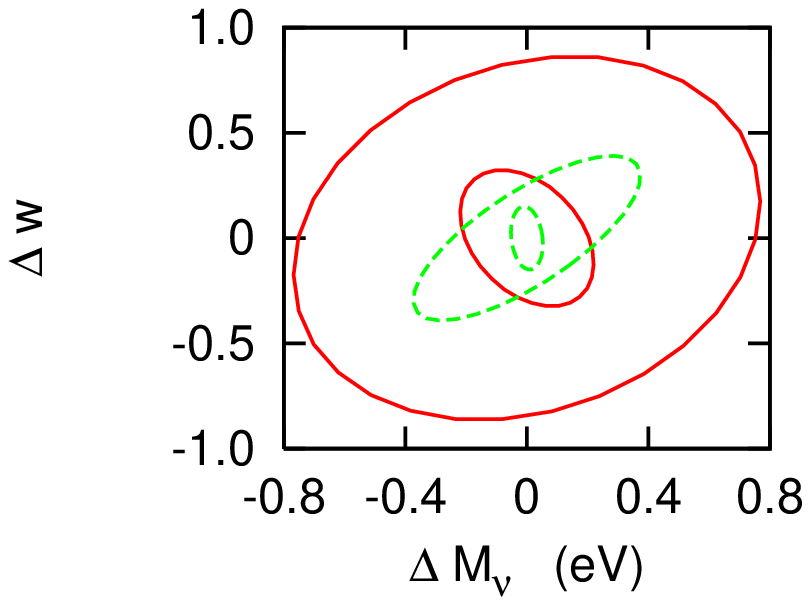}
\includegraphics[width=.23\textwidth]{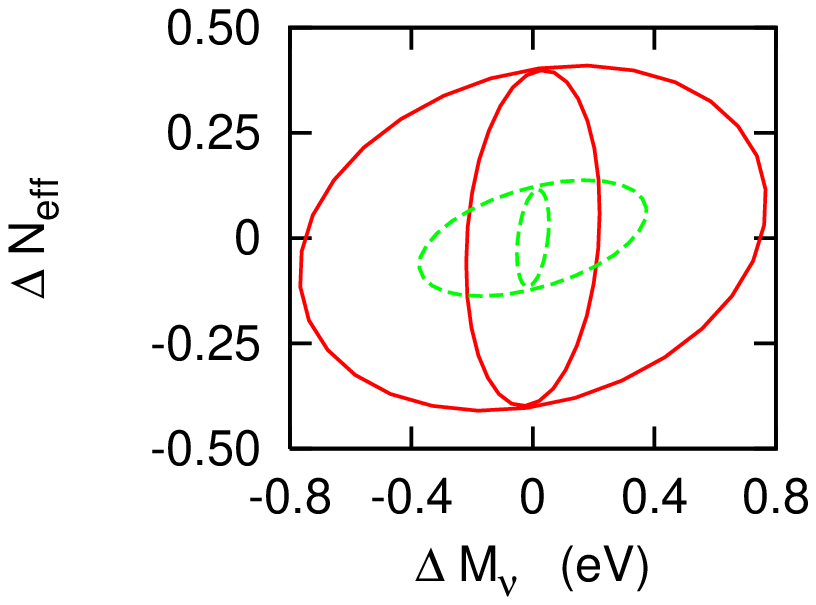}
\caption{\label{fig_ellipses} 1-$\sigma$ confidence limits on the
pairs ($M_{\nu}$, $\theta_i$), for each parameter $\theta_i$ in our
eleven-dimensional model. The red solid (green dashed) contours
are those expected for Planck (Inflation Probe). For each case,
the smaller (larger) ellipse corresponds to the forecasts
with (without) lensing extraction.}
\end{figure}

Finally, the version of the Inflation Probe satellite that we consider
here is able to reach $\sigma(M_{\nu}) = 0.035$ eV both in the eight
and eleven parameter cases. Note that when we take instead the CMBpol
specifications of Ref.\ \cite{Kaplinghat:2003bh}, we exactly reproduce
their forecast $\sigma(M_{\nu}) = 0.04$ eV (derived for an
intermediate case with ten parameters). It is interesting to see that
even with such a precise experiment, the results are robust against
foreground contamination, since in absence of any cleaning the
forecast error increases only by $15\%$.

We show in Fig.\ \ref{fig_ellipses} the correlation between $M_{\nu}$
and each free parameter of the eleven-dimensional model, in the cases
of Planck and Inflation Probe, with and without lensing extraction. In
the parameter basis used in this work, the neutrino mass appears to be
mainly degenerate with $\omega_m$, and the lensing extraction process
removes most of this degeneracy.

\section{Future sensitivities to the neutrino mass splitting}

In principle, the LSS power spectrum is not sensitive only to the
total mass $M_{\nu}$, but also to the way in which the mass is
distributed among the three neutrino states.  The reason is twofold:
the amount by which the gravitational collapse of matter perturbation
is slowed down by neutrinos on small scales depends on the time of the
non-relativistic transition for each eigenstate, i.e. on the
individual masses; and the characteristic scales at which the
free-streaming effect of each neutrino family is imprinted in the
power spectrum depends on the value of the wavelengths crossing the
Hubble radius at the time of each non-relativistic transition,
i.e. again on the individual masses.

The neutrino masses are differently distributed among the three states
in the two possible mass schemes, or hierarchies, as shown e.g.\ in
Fig.\ 1 of \cite{Lesgourgues:2004ps}. For a total mass $M_{\nu}$
larger than $0.2-0.3$ eV all neutrino states approximately share the
same mass $m_0=M_\nu/3$, in the so-called degenerate region. Instead,
for smaller $M_{\nu}$ the splitting between the individual masses is
more visible, and for the minimum values of $M_{\nu}$ one finds that
in the Normal Hierarchy case (NH) there is only one neutrino state
with significant mass, or two degenerate states in the Inverted
Hierarchy case (IH). In general, for a given $M_{\nu}$ one can
calculate the difference between the matter power spectrum in the two
cases, as has been computed numerically in Ref.\
\cite{Lesgourgues:2004ps}.

We would like to study whether the lensing power spectrum derived from
a very precise CMB experiment like Inflation Probe would be able to
discriminate between the two models. For this purpose, we take the
eight parameter model of section \ref{sec:results} and complete it
with a ninth parameter: the number of massive neutrinos $N_{\nu}^{\rm
massive}$, which could be equal to 1, 2 or 3 (the remaining species
are assumed to be exactly massless).  In a NH scenario with
$M_{\nu}>0.1$~eV, the mass of the third neutrino is not completely
negligible: so, we expect the difference between our simplified
scenario with $N_{\nu}^{\rm massive}=1$ and that with $N_{\nu}^{\rm
massive}=2$ to be more pronounced than the difference between
realistic NH and IH scenarios (assuming the same total mass $M_{\nu}$
in all models). This statement is confirmed by the numerical results
of Ref.\ \cite{Lesgourgues:2004ps}. So, if we could show that an
experiment like Inflation Probe will be unable to differentiate
between the sketchy $N_{\nu}^{\rm massive}=1$ and $N_{\nu}^{\rm
massive}=2$ models, we would conclude that {\it a fortiori} it will
not discriminate between the NH and IH scenarios.

We repeated the computations of section \ref{sec:results} with a ninth
free parameter $N_{\nu}^{\rm massive}$ with fiducial value
$N_{\nu}^{\rm massive}=1$. Note that the parameter $N_{\nu}^{\rm
massive}$ should not be confused with the total effective neutrino
number $N_{\rm eff}$, which was a free parameter in the last section,
and remains fixed to $N_{\rm eff}=3$ in the present one.  We found
for Inflation Probe -- including lensing extraction and assuming
perfect foreground cleaning-- a one-sigma error
$\sigma\left(N_{\nu}^{\rm massive}\right)=2.8$.  We conclude that the
experiments and techniques discussed in the present paper are far from
sufficient for discriminating between the NH and IH scenarios. In any
case, as shown in Ref.\ \cite{Lesgourgues:2004ps}, future results on
the total neutrino mass from very precise cosmological data should be
interpreted in a slightly different way for the NH and IH cases.

\section{Conclusions}

We have studied the ability of future CMB experiments to measure the
power spectrum of large scale structure, using some quadratic
estimators of the weak lensing deflection field. We inferred the
sensitivity of these experiments to the non-zero neutrino masses
indicated by neutrino oscillation data. Our aim was to extend the
pioneering paper by Kaplinghat, Knox \& Song \cite{Kaplinghat:2003bh}
by further investigating several directions.

First, we based our analysis on the following list of forthcoming CMB
experiments (either operational, approved or still in project):
BICEP, QUaD, BRAIN, ClOVER and {\sc Planck}, SAMPAN and Inflation
Probe, taking into account their detailed characteristics.  We found
that even before {\sc Planck}, ground-based experiments should succeed
in extracting the lensing map with good precision, and could then
significantly improve the bounds on neutrino masses. We also found
that the SAMPAN mini-satellite project would be able to reduce the
Planck error $\sigma(M_{\nu})$ from approximately $0.15$ eV to $0.10$
eV. Finally, the hypothetical version of Inflation Probe that we
considered would reach a spectacular sensitivity of
$\sigma(M_{\nu})\sim 0.035$ eV. 

We also tried to discuss two questions raised by the analysis of Ref.\
\cite{Kaplinghat:2003bh}: first, is it really accurate to base the
Fisher matrix computation on perfectly delensed maps on the one hand,
and on the reconstructed lensing map on the other?  Second, is it
realistic to estimate the noise variance of the lensing quadratic
estimators without taking into account any residual foreground
contamination? Our answer to these two questions is positive: we did
not provide an exact treatment of these very technical issues, but we
tried to systematically bracket the results between two
over-optimistic and over-pessimistic assumptions, and concluded that
the error forecast method of Ref.~\cite{Kaplinghat:2003bh} is robust.

Finally, we investigated the issue of parameter degeneracies involving
the neutrino mass, by comparing the results in a simpler model than
that of Ref.\ \cite{Kaplinghat:2003bh} with those in a more
complicated one. Our extended cosmological model allows for a scalar
tilt running, a dark energy equation of state parameter $w \neq -1$,
and extra degrees of freedom parametrized by the effective number of
massless neutrinos $N_{\rm eff}$. These extra parameters were not
chosen randomly.  The tilt running was shown to be slightly degenerate
with the neutrino mass in an analysis involving current CMB and LSS
data \cite{Seljak:2004xh}.  The same holds for the equation of state
of dark energy \cite{Hannestad:2005gj} and for the effective number of
massless neutrinos \cite{Hannestad:2003ye,Crotty:2004gm}. However, our
results indicate that future CMB experiments will be able to resolve
these degeneracies, since we do not find significant discrepancies
between the neutrino mass errors obtained for our two cosmological
models.

Fortunately, CMB lensing extraction should be regarded as only one of
the most promising tools for measuring the absolute neutrino mass with
cosmology. It could be combined with future data from tomographic
galaxy cosmic shear surveys, which will be very sensitive to neutrino
masses \cite{Song:2003gg}.  The cross-correlation of LSS information
with CMB temperature anisotropies could also reveal very useful for
the purpose of measuring $M_{\nu}$ \cite{Ichikawa:2005hi}. In the
method employed in the present paper, the correlation between
temperature and lensing (the $Td$ term) is already taken into account,
but it affects the final results only marginally.  More interesting
should be the cross-correlation of future data from large cosmic shear
surveys with that from CMB anisotropies.

In conclusion, our results show that there are good perspectives to
detect non-zero neutrino masses using future CMB lensing data, since
even in the less favorable case of the smallest $M_{\nu}\simeq 0.05$
eV in the NH mass scheme the Inflation Probe experiment alone could
make a marginal detection (between the one and two sigma
levels). Obviously the sensitivity is enhanced for larger values of
$M_{\nu}$, in particular for the mass degenerate and quasi-degenerate
regions but also for the minimum of $M_{\nu} \simeq 0.1$ eV in the IH
case.  The information on $M_\nu$ from analyses of cosmological data
will be complementary (and vice versa) to the efforts in terrestrial
projects such as tritium beta decay and neutrinoless double beta decay
experiments. Of course any positive result on the absolute neutrino
mass scale will be a very important input for theoretical models of
particle physics beyond the Standard Model.

\section*{Acknowledgments}
We would like to thank Karim Benabed, Simon Prunet and Jonathan Rocher
for extremely useful discussions.  This work was supported by a
MEC-IN2P3 agreement. SP was supported by the Spanish grants
BFM2002-00345 and GV/05/017 of Generalitat Valenciana, as well as by a
Ram\'{o}n y Cajal contract of MEC.

\end{document}